\documentclass[preprint]{IEEEtran}

\usepackage{cite}
\usepackage{graphicx}
\usepackage[noend]{algorithm,algpseudocode}
\usepackage{algorithmicx}

\usepackage{amssymb}
\usepackage{mathrsfs}
\usepackage{amsmath}

\usepackage[sans]{dsfont}
\usepackage{stmaryrd}
\usepackage{pifont}
\usepackage{wasysym}

\usepackage{array}
\usepackage{url}
\usepackage[usenames]{color}
\usepackage[tight,footnotesize]{subfigure}
\usepackage{multirow}
\usepackage{enumitem}
\algnewcommand{\LeftComment}[1]{\Statex \(\triangleright\) #1}

\usepackage{flushend}

\newtheorem{lemma}{Lemma}

\newtheorem{innercustomthm}{Theorem}
\newenvironment{customthm}[1]
{\renewcommand\theinnercustomthm{#1}\innercustomthm}
{\endinnercustomthm}

\begin{document}

\title{Drive Mode Optimization and Path Planning for Plug-in Hybrid Electric Vehicles}

\author{Chi-Kin Chau, Khaled Elbassioni and Chien-Ming Tseng
\thanks{C.-K. Chau, K. Elbassioni, and C.-M. Tseng are with the Department of EECS at Masdar Institute of Science and Technology, UAE (e-mail: \{ckchau, kelbassioni, ctseng\}@masdar.ac.ae).}
\thanks{This paper appears in IEEE Transactions on Intelligent Transportation Systems (DOI:10.1109/TITS.2017.2691606).}
}

\pagestyle{plain}
\thispagestyle{plain}

\maketitle

\begin{abstract}
Drive modes are driver-selectable pre-set configurations of powertrain and certain vehicle parameters. Plug-in hybrid electric vehicles (PHEVs) typically feature special options of drive modes that can affect the hybrid energy source management system, for example, electric vehicle (EV) mode (that draws fully on battery) and charge sustaining (CS) mode (that utilizes internal combustion engine to charge battery while propelling the vehicle). This paper studies an optimization problem to enable the driver to select the appropriate drive modes for fuel minimization. We develop optimization algorithms that optimize the decisions of drive modes based on trip information, and integrated with path planning to find an optimal path, considering intermediate filling and charging stations. We further provide an online algorithm that is based on the revealed trip information. We evaluate our algorithms empirically on a Chevrolet Volt, which shows significant fuel savings.
\end{abstract}


\section{Introduction}

Modern vehicles are provided with a plethora of configuration options. {\em Drive modes} are a set of pre-set profiles of configurations of powertrain and other vehicle parameters that are selectable by drivers during driving. For example, Sport mode maximizes the engine performance by allowing larger horsepower, whereas ECO mode suppresses the vehicle performance by constraining acceleration and throttle response.

In this paper, we consider the drive modes specifically in plug-in hybrid electric vehicles (PHEVs). PHEVs are equipped with rechargeable batteries and electrical machines (which double as electric motors and generators), as well as conventional internal combustion engines. PHEVs are benefited by the convenience of fuel refilling and cheap electrical energy. In addition to regenerative braking and electric-motor-assisted engine stop/start, PHEVs can harness the diversity of energy efficiency of electric motors and combustion engines by optimizing the hybrid energy sources.

There are special drive modes in PHEVs that can affect the hybrid energy source management system. For example, {\em Electric Vehicle} (EV) mode allows the PHEV to draw solely on battery without relying on internal combustion engine, and {\em Charge Sustaining} (CS) mode utilizes internal combustion engine to charge battery and propel the PHEV simultaneously.

Many prior results \cite{mapelli2009ecshev,theo2009rbemhv,kessels2008oemhev} in energy management consider the internal optimization processes of PHEVs, which assume complete controls of all system components of a PHEV (see Sec.~\ref{sec:related}). As a departure from prior work, this paper focuses on a driver-centric approach to let the driver to select the appropriate drive modes while driving.

In this paper, we consider general optimization problems of drive modes for fuel minimization in the following settings:
\begin{itemize}

\item {\bf Route-based Drive Mode Optimization}:
Given the trip information for a particular route (e.g., a forecast of vehicle speed profile), we find an optimal solution of drive mode decisions for each segment of the trip.

\item {\bf Online Drive Mode Optimization}:
The decisions of drive modes are made in an online fashion, based on only the revealed trip information over time.

\item {\bf Integrated Drive Mode Optimization and Path Planning}:
Given the source and destination of a trip, we find an optimal path with a solution of drive mode decisions, taking into account various fuel prices at intermediate filling stations and the availability of battery charging.

\end{itemize}

In Sec.~\ref{sec:model}, we formulate the preceding problem by an integer programming problem, which captures several practical aspects of PHEVs (e.g., multi-mode transmission, and vehicle speed dependency in combustion engine management). In Sec.~\ref{sec:solution}, we devise effective algorithms for solving these drive mode optimization problems. We also provide a fast approximation algorithm that can scale with large problem sizes.
To demonstrate the practical value of our results, we evaluate our algorithms empirically on a Chevrolet Volt in Sec.~\ref{sec:eval}. Validated by real-world data, we observe that our system can provide a significant improvement in fuel savings. We also discuss several practical issues in Sec.~\ref{sec:disc}.

The future vehicle platforms will be likely to support third-party applications. Thus, our system can be loaded as a software application on the vehicle for automatic drive mode decisions, even without explicitly relying on drivers' inputs.

\section{Background} \label{sec:background}

The powertrain mechanics of PHEVs can be classified as series hybrids, parallel hybrids, and series/parallel hybrids. In series hybrids, the internal combustion engine is always connected to the generator to charge the battery. The drivetrain is only powered directly by electric motor. Once the state-of-charge (SoC) of battery becomes low, the internal combustion engine will start to charge the battery.
In parallel hybrids, the combustion engine and electric motors can operate in tandem to power the drivetrain.
There is a power split system to combine the parallel power sources, and a possible clutch to enable the combustion engine to charge battery and propel the vehicle simultaneously.
In series/parallel hybrids, a combination of planetary gear trains allows flexible power split between the combustion engine and a number of electric motors.

Despite the differences of powertrain mechanics, the internal operations of PHEVs are often transparent to drivers. There are automatic systems to manage the transmission gear, powertrain and hybrid energy sources. Given the steering and pedal control status, the automatic system controls the torque, rpm of combustion engine\footnote{In practical PHEV systems, the rpm of combustion engine is normally related to the vehicle speed, even for series hybrids in which the combustion engine is not directly connected to the drivetrain. Apparently, there is a safety hazard for the combustion engine operating in a high speed, when the vehicle is stationary. Also, production PHEVs often use a variable number of electric motors/generators conditional on the vehicle speed. In high speed, more than one electric motor may be used.}, transmission gears, output power of battery, etc., to match the load of drivetrain accordingly \cite{mapelli2009ecshev,theo2009rbemhv}.
Although the low-level mechanics are not controllable by drivers, typical vehicles are usually customizable by setting certain high-level drive modes. We survey the available drive modes in several production PHEV models that can affect the behavior of hybrid energy management system in Table~\ref{tab:modes}.

\begin{table}[!htb]
\centering
  \caption{Examples of drive modes in production PHEVs.} \label{tab:modes}
\begin{tabular}{@{}p{66pt}@{ }|@{}c@{}|@{ }p{140pt}@{}}
\hline
\hline
    \quad Vehicle Model & Drive Mode & \qquad \qquad \qquad Description \\
\hline
	\multirow{7}{*}{\parbox{66pt}{\mbox{Chevrolet Volt} (model 2011-2015)}} 	
	& \multirow{3}{*}{Normal}  & Draw only on battery until SoC drops to 22\%. Then, use combustion engine to charge battery and propel vehicle.  \\ \cline{2-3}
	& \multirow{2}{*}{Mountain} & Same as Normal mode, but draw only on battery until SoC drops to 45\%. \\  \cline{2-3}
	& \multirow{2}{*}{Hold} & Use combustion engine to maintain the current SoC. \\
\hline	
	\multirow{4}{*}{\parbox{66pt}{Toyota Prius Plug-in (model 2012-2015)}} 	
	& \multirow{2}{*}{Normal} & Use both combustion engine and electric motor to propel vehicle. \\ \cline{2-3}
	& \multirow{2}{*}{EV} & Always draw on battery, if there is sufficient SoC and within EV mode speed limit. \\
\hline
	\multirow{6}{*}{\parbox{66pt}{Ford Fusion Energi (model 2014-2015)}} 	
	& \multirow{2}{*}{EV Now} & Draw on battery entirely, if there is sufficient SoC. \\ \cline{2-3}
	& \multirow{2}{*}{EV Auto} & Use both combustion engine and electric motor, depending on vehicle speed. \\ \cline{2-3}
	& \multirow{2}{*}{EV Later} & Conserve battery by mostly using combustion engine, which also charge battery. \\
\hline \hline
  \end{tabular}
\end{table}

Since there are a variety of model-specific drive modes, this paper aims to present a study as general as possible and to be extensible to future models. Rather than considering model-specific drive modes, we consider four {\em generic drive modes}: (1) {\em Electric Vehicle} (EV) mode that draws solely on battery, (2) {\em Combustion Engine} (CE) mode that relies solely on internal combustion engine, (3) {\em Charge Sustaining} (CS) mode that utilizes internal combustion engine to charge battery and propel the vehicle, and (4) {\em Aggregate Power} (AP) mode that combines both electric motor and internal combustion engine to propel the vehicle. In Sec.~\ref{sec:model}, we formulate a general optimization problem of generic drive mode decisions. We remark that our problem is sufficiently general to capture a variety of existing PHEV models. We will discuss the mapping from the model-specific drive modes to generic drive modes, and validate our model for Chevrolet Volt empirically.

\section{Related Work} \label{sec:related}

There is a body of work about optimizing energy management systems for PHEVs. For example, \cite{mapelli2009ecshev} uses heuristic control strategy to optimize energy consumption for given torque and speed. A similar concept relying on rule-based management policies has been presented in \cite{theo2009rbemhv}.
\cite{qiuming2008tbopm} considers continuous-time optimization control of hybrid energy sources. Some studies focus on sub-optimal solutions that can be computed faster than dynamic programming \cite{cong2014apmpphev}\cite{namwook2012opteqfchev}. 
These prior results often assume complete controls of internal energy management system in PHEVs, and are mostly based on simulations. Our work considers limited control by only selecting the drive modes available in the PHEVs, and we evaluate the results on a real-world PHEV. The models of PHEVs and experimental validations were studied in \cite{mapelli2012phevmdlpty, sciarretta2014bechmarkphev}.

This papers utilizes the estimated information of vehicles. Participatory sensing can provide data with good geographic penetrations \cite{cloudthink15}. Our previous work employs participatory sensing for distance-to-empty prediction \cite{cmtseng2015pardte,cmtseng2017dte,cmtseng2016privacy}.

There are prior papers about online energy management strategies for PHEVs. For example, \cite{kessels2008oemhev} utilizes nonlinear optimization for parallel and serial HEVs. However, this paper considers competitive online algorithms \cite{allan1998occa} that can provide proven worst-case guarantees from the offline optimal solutions.
The path planning problem considering various gas prices at filling stations has been studied in\cite{samir2011fillgas}.
However, this paper considers a more general path planning problem with energy management strategies for PHEVs.
This paper extends substantially the preliminary results in \cite{ckchau2016phevopt}. We now provide approximation solution to path planning and extend the online algorithm to consider four generic drive modes.

\section{Model and Problem Formulation} \label{sec:model}

\begin{table}[hbt] %
\caption{Key Notations}\label{tab1} %
\centering
\begin{tabular}
[l]{@{}c@{}|@{}l@{}}
\hline
\textbf{Notation} &  \textbf{\ Definition}\\
\hline
$B_t$ & The state-of-charge of the battery at time $t$.\\
$G_t$ & The fuel tank level at time $t$.\\
$P_t$ & The power of drivetrain of PHEV at time $t$.\\
$s_t$ & The power from battery to electric motor at time $t$.\\
$r_t$ & The power from generator to charge the battery at time $t$.\\
$\eta^{\rm r}_t$,$\eta^{\rm d}_t$ & The charging and discharging efficiency coefficients at time $t$.\\
$Q_t$ & The output power from combustion engine at time $t$.\\
$F(\cdot)$ & A function maps the output power to the required amount of fuel.\\
$f$ & The per-unit cost of $F(\cdot)$, $f(Q) \triangleq \tfrac{F(Q)}{Q}$.\\
$u_t$ & The power from engine to charge the battery at time $t$.\\
$\eta^{\rm e}_t$ & The charging efficiency coefficient by combustion engine at time $t$.\\
$C_t$ & The maximum available power to charge battery \\
 & from combustion engine at time $t$.\\
$\beta_t$ & The maximum portion of power contributed by electric motor at time $t$.\\
\hline
\end{tabular} %
\end{table}

Our goal is to develop a systematic study for drive mode decisions, based on a generic vehicle model characterized by parameters using measurements or standard vehicle information. We consider a semi-blackbox model of PHEV that is abstracted away from the underlying vehicle control systems. A table of key notations is given in Table~\ref{tab1}.

This paper considers a discrete-time setting from time slot $t=1$ to $t=T$, where the inputs within one time slot are assumed to be quasi-static. Let $G_t$ be the fuel tank level and $B_t$ be the SoC of the PHEV at time $t$. $G_0$ and $B_0$ are the initial fuel tank level and SoC respectively.

Define the driving profile be $(v_t, \alpha_t)_{t=1}^T$, where $v_t$ is the vehicle speed and $\alpha_t$ is the gradient of road at time $t$. The driving profile can be obtained by prediction using historic data, or crowd-sourced data collection \cite{ganti2010greengps,cloudthink15,cmtseng2015pardte,cmtseng2017dte,cmtseng2016privacy}. Note that $v_t$ is non-negative for all $t$. We assume that the energy consumption of PHEV is solely characterized by the driving profile, for example, under moderate weather and traffic conditions.
Let the acceleration be $a_t \triangleq v_t - v_{t\mbox{-}1}$.
The load of drivetrain of a generic vehicle \cite{eugene2013rtbattery,karin2012eeroutealg} is given~by
\begin{equation}
P_t = \textstyle \frac{\rho_{\rm a}k_{\rm d} A_{\rm f} v_t^3}{2}  + {\sf m} {\sf g} \sin(\alpha_t) v_t + {\sf m} {\sf g} k_{\rm r} v_t + {\sf m} v_t a_t + {\sf c}_0 \label{eqn:vehiclemodel} 
\end{equation}
where ${\sf m}$ is the vehicle weight, ${\sf g}$ is the gravitational constant, $\rho_{\rm a}$ is the density of air, $A_{\rm f}$ is the frontal area of the vehicle, $k_{\rm d}$ is aerodynamic drag coefficient of the vehicle, $k_{\rm r}$ is the rolling friction coefficient, and ${\sf c}_0$ is the default load (e.g., due to air-conditioning). These parameters can be obtained from standard vehicle information or simple measurement.

Note that $P_t$ can be positive or negative (possibly due to negative $a_t$). Let $P^+_t = \max\{P_t, 0\}$ and $P^-_t = - \min\{P_t, 0\}$. $P^-_t$ represents the power captured by regenerative breaking.

\begin{figure}[htb!]
	\centering  
    \includegraphics[scale = 0.35]{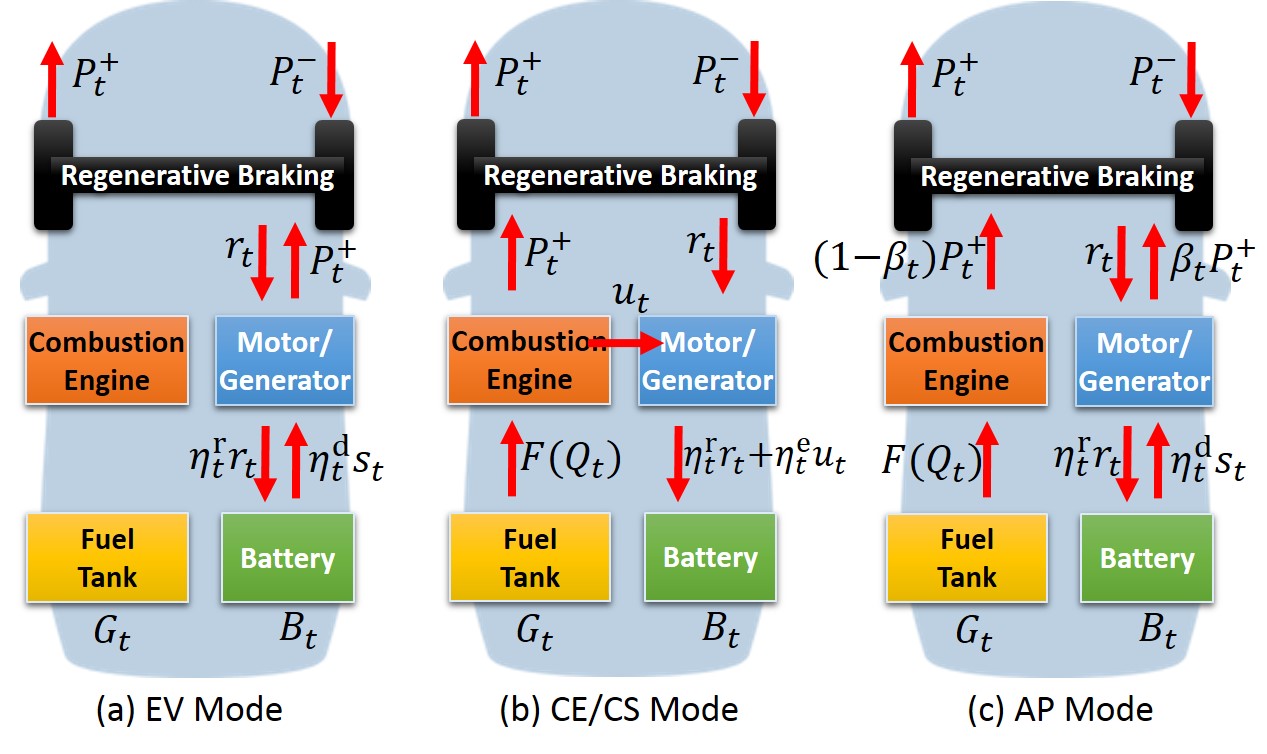} 
    \caption{System models of four generic drive modes.}\label{fig:modes} 
\end{figure}

In the following subsections, we describe the four generic drive modes (EV, CE, CS, AP modes), as illustrated in Fig.~\ref{fig:modes}. These generic drive modes provide abstract representations of the model-specific drive modes in PHEVs.

\subsection{Electric Vehicle (EV) Mode}

In EV mode (also called charge depleting mode), the PHEV is only powered by battery, which is also charged by regenerative braking when decelerating or stopping. Let $[\underline{B},\overline{B}]$ be the allowable range of SoC to operate in EV mode.
Let $s_t$ be the power from battery to electric motor (when $P^+_t \ge 0$), and $r_t$ be the power from generator to charge the battery (when $P^-_t \ge 0$).
If $\underline{B} \le B_{t\mbox{-}1} \le \overline{B}$, then the SoC is given by 
\begin{equation}
B_t = B_{t\mbox{-}1} + \eta^{\rm r}_t   r_t - \eta^{\rm d}_t   s_t \label{eqn:evmodebattery} 
\end{equation}
subject to $\underline{B} \le B_{t} \le \overline{B}$, $r_t \le P^-_t$ and $s_t = P^+_t$.
Parameters $\eta^{\rm r}_t\le 1$ and $\eta^{\rm d}_t\ge 1$ denote the charging and discharging efficiency coefficients. Note that $\eta^{\rm r}_t$ and $\eta^{\rm d}_t$ are time-variable\footnote{The efficiency coefficients are often assumed time-invariant in previous work (e.g., \cite{mapelli2009ecshev,theo2009rbemhv,kessels2008oemhev}).}, because there may be a variable number of generators/motors to be utilized in the PHEV, depending on the driving profile (as observed in production PHEVs such as Chevrolet Volt).

Note that regenerative braking incurs no fuel cost. Hence, Eqn.~(\ref{eqn:evmodebattery}) and the constraints are equivalent to setting 
\begin{equation}
s_t = \min\{P^+_t, \textstyle \frac{B_{t\mbox{-}1} - \underline{B}}{\eta^{\rm d}_t } \},\
r_t = \min\{P^-_t, \textstyle \frac{\overline{B} - B_{t\mbox{-}1}}{\eta^{\rm r}_t } \} 
\end{equation}

\subsection{Combustion Engine (CE) Mode}

In CE mode, the PHEV is only powered by internal combustion engine. Let the output power from combustion engine at time $t$ be $Q_t$.
The fuel tank level is given by
\begin{equation}
G_t = G_{t\mbox{-}1} - F(Q_t), \mbox{\ where\ } Q_t = P^+_t
\end{equation}
subject to $G_t \ge 0$. $F(\cdot)$ is an increasing convex function that maps the output power to the required amount of fuel.

We also allow the battery to be charged by regenerative braking, if possible. The SoC is given by
\begin{equation}
B_t = B_{t\mbox{-}1} + \eta^{\rm r}_t r_t
\end{equation}
subject to $B_{t} \le \overline{B}$ and $r_t \le P^-_t$. This implies that $r_t = \min\{P^-_t,$ $\frac{\overline{B} - B_{t\mbox{-}1}}{\eta^{\rm r}_t } \}$.

\subsection{Charge Sustaining (CS) Mode}

In CS mode, the internal combustion engine is used to propel the vehicle and charge the battery simultaneously.
Let $u_t$ be the power from engine to charge the battery at time $t$.
The fuel tank level is given by  
\begin{equation}
G_t = G_{t\mbox{-}1} - F(Q_t), \mbox{\ where\ } Q_t = P^+_t + u_t \label{eqn:csmodefuel} 
\end{equation}
subject to $G_t \ge 0$. The SoC is given by  
\begin{equation}
B_t = B_{t\mbox{-}1} + \eta^{\rm r}_t r_t + \eta^{\rm e}_t u_t \label{eqn:csmodebattery} 
\end{equation}
subject to $B_{t} \le \overline{B}$, $r_t \le P^-_t$, and $u_t \le C_t$.
$\eta^{\rm e}_t\le 1$ denotes the charging efficiency coefficient by combustion engine. In this paper, we allow possibly different generators used by regenerative braking and combustion engine (as observed in Chevrolet Volt). $C_t$ is the maximum available power to charge battery from combustion engine at time $t$.

Note that practical control system in PHEVs (e.g., Chevrolet Volt) may set the rpm of combustion engine related to vehicle speed. $C_t$ captures the limitation of available power from combustion engine depending on vehicle speed at time $t$.

We assume that the energy management system in PHEV attempts to charge battery up to $C_t$ whenever possible. Then, Eqns.~(\ref{eqn:csmodefuel})-(\ref{eqn:csmodebattery}) and the constraints are equivalent to setting 
$$ \hspace{-11pt}
r_t = \min\{P^-_t, \textstyle \frac{\overline{B} - B_{t\mbox{-}1}}{\eta^{\rm r}_t } \},\
u_t = \min\{C_t, \textstyle \frac{\overline{B} - B_{t\mbox{-}1} - \eta^{\rm r}_t r_t}{\eta^{\rm e}_t } \} 
$$

\subsection{Aggregate Power (AP) Mode}

In AP mode, the PHEV is propelled by internal combustion engine and electric motor (that is powered by battery) together.
The SoC is given by  
\begin{equation}
B_t = B_{t\mbox{-}1} + \eta^{\rm r}_t r_t - \eta^{\rm d}_t s_t \label{eqn:apmodebattery} 
\end{equation}
subject to $\underline{B} \le B_{t} \le \overline{B}$, $r_t \le P^-_t$ and $s_t \le \beta_t P^+_t$. $\beta_t \le 1$ is the maximum portion of power contributed by electric motor to drivetrain. In this paper, we allow a variable number of generators/motors to be utilized in the PHEV, conditional on the driving profile, which can induce a variable portion of power split by electric motor and combustion engine over time.
The fuel tank level is given by  
\begin{equation}
G_t = G_{t\mbox{-}1} - F(Q_t), \mbox{\ where\ } Q_t = P^+_t - s_t  \label{eqn:apmodefuel} 
\end{equation}
subject to $G_t \ge 0$. We assume that the energy management system in PHEV attempts to use electric motor to power the drivetrain by $\beta_t P^+_t$ whenever possible. Then, Eqns.~(\ref{eqn:apmodebattery})-(\ref{eqn:apmodefuel}) and the constraints are equivalent to setting 
$$ \hspace{-5pt}
s_t = \min\{\beta_t P^+_t, \textstyle \frac{B_{t\mbox{-}1} - \underline{B}}{\eta^{\rm d}_t } \},\
r_t = \min\{P^-_t, \textstyle \frac{\overline{B} - B_{t\mbox{-}1}}{\eta^{\rm r}_t } \} 
$$

Note that our model is abstracted away from the underlying mechanics, like automatic transmission, powertrain control, etc. But the parameters $\eta^{\rm r}_t, \eta^{\rm d}_t, \eta^{\rm e}_t, C_t, \beta_t$ are sufficiently general to capture the effects of the underlying mechanics.

\subsection{Mapping to Generic Drive Modes} \label{sec:mapping}

We next discuss the mapping from model-specific drive modes to generic drive modes. Chevrolet Volt has four drive modes: Normal, Sport, Mountain and Hold. We consider EV and CS modes only, and SoC lies in the ranges $[22\%, 45\%]$. To trigger EV mode, one can enter Normal mode. To trigger CS mode, one can enter Mountain mode. If SoC is higher than $45\%$, only EV mode is allowed. If SoC is lower than $22\%$, only CS mode is allowed.

\begin{figure*}[htp!]
\begin{minipage}[t]{0.5\textwidth}
\begin{align}
(\textsc{DMOP}) & \qquad \displaystyle \min_{(x_t)_{t=1}^T} \  {\tt Cost}(T) = G_0 - G_T \notag\\
\mbox{subject to} & \mbox{\ for all\ } t \in [1, T], \notag\\
 & G_t = G_{t\mbox{-}1} - F(Q_t),\ G_t \ge 0, \label{con:dmop-f}\\
 & Q_t = (1 - x^{\rm ev}_t)P^+_t + x^{\rm cs}_t u_t - x^{\rm ap}_t s_t, \label{con:dmop-q}\\
 & B_t = B_{t\mbox{-}1} + \eta^{\rm r}_t r_t+ \eta^{\rm e}_t u_t - \eta^{\rm d}_t s_t, \label{con:dmop-b} \\
 & \underline{B} \le  B_{t} \le \overline{B}, \label{con:dmop-bb}\\
 & r_t = \min\{P^-_t, \textstyle \frac{\overline{B} - B_{t\mbox{-}1}}{\eta^{\rm r}_t } \}, \label{con:dmop-r}\\
 & s_t = \min\{x^{\rm ev}_t P^+_t + x^{\rm ap}_t \beta_t P^+_t, \textstyle \frac{B_{t\mbox{-}1} - \underline{B}}{\eta^{\rm d}_t } \}, \label{con:dmop-s}\\
 & u_t = \min\{x^{\rm cs}_t C_t, \textstyle \frac{\overline{B} - B_{t\mbox{-}1} - \eta^{\rm r}_t r_t}{\eta^{\rm e}_t } \}, \label{con:dmop-u}\\
 & x^{\rm ev}_t + x^{\rm ce}_t + x^{\rm cs}_t + x^{\rm ap}_t = 1, \label{con:dmop-x}\\
 & x^{\rm ev}_t, \ x^{\rm ce}_t, \ x^{\rm cs}_t, \ x^{\rm ap}_t \in \{0, 1\} \label{con:dmop-x0}
\end{align}
\end{minipage}
\begin{minipage}[t]{0.5\textwidth}
\begin{align}
(\textsc{cDMOP}) & \ \displaystyle \min_{(x_t, r_t, s_t, u_t)_{t=1}^T}  {\tt Cost}(T) = G_0 - G_T \notag\\
\mbox{subject to} & \mbox{\ for all\ } t \in [1, T], \notag\\
 & G_t = G_{t\mbox{-}1} - F(Q_t),\  G_t \ge 0, \label{con:cdmop-f}\\
 & Q_t = (1 - x^{\rm ev}_t)P^+_t + x^{\rm cs}_t u_t - x^{\rm ap}_t s_t,  \label{con:cdmop-q}\\
 & B_t = B_{t\mbox{-}1} + \eta^{\rm r}_t r_t + \eta^{\rm e}_t u_t - \eta^{\rm d}_t s_t, \label{con:cdmop-b} \\
 & \underline{B} \le  B_{t} \le \overline{B}, \label{con:cdmop-bb}\\
 & r_t \le P^-_t, \label{con:cdmop-r}\\
 & x^{\rm ev}_t P^+_t \le s_t \le x^{\rm ev}_t P^+_t + x^{\rm ap}_t \beta_t P^+_t, \label{con:cdmop-s}\\
 & u_t \le x^{\rm cs}_t C_t, \label{con:cdmop-u}\\
 & r_t, \ s_t, \ u_t \ge 0, \label{con:cdmop-usr} \\
 & x^{\rm ev}_t + x^{\rm ce}_t + x^{\rm cs}_t + x^{\rm ap}_t = 1, \label{con:cdmop-x}\\
 & 0 \le x^{\rm ev}_t, \ x^{\rm ce}_t, \ x^{\rm cs}_t, \ x^{\rm ap}_t \le 1 \label{con:cdmop-x0}
\end{align}
\end{minipage} 
\end{figure*}

Toyota Prius Plug-in provides four drive modes: ECO, Normal, Power and EV. EV mode is present. Normal mode may be mapped to CS or CE modes, whereas Power mode may be mapped to AP or CE modes.
Ford Fusion Energi offers three drive modes: EV Now, EV Auto and EV Later. EV mode is present. EV Auto mode may be mapped to AP mode, whereas EV later mode may be mapped to CS mode.

The mapping of model-specific drive modes to generic drive modes can be validated by empirical studies. In this paper, we validate our model particularly for Chevrolet Volt in Sec.~\ref{sec:eval}.

\subsection{Drive Mode Optimization Problem}

Let $x^{\rm ev}_t$, $x^{\rm ce}_t$, $x^{\rm cs}_t$, $x^{\rm ap}_t$ be the binary decision variables indicating if EV, CE, CS, and AP modes are enabled at time $t$, respectively. Let $x_t \triangleq (x^{\rm ev}_t, x^{\rm ce}_t, x^{\rm cs}_t, x^{\rm ap}_t)$. We formulate a fuel minimization problem of drive mode decisions by integer programming problem (\textsc{DMOP}), given a driving profile $(v_t, \alpha_t)_{t=1}^T$, initial SoC $B_0$ and fuel tank level $G_0$. The objective of \textsc{DMOP} is to minimize the total cost: ${\tt Cost}(T) \triangleq G_0 - G_T$.

Note that \textsc{DMOP} does not always have a feasible solution, for example, when there is insufficient fuel. Also, if a drive mode is not present, we can disable a certain drive mode in \textsc{DMOP} by adding an additional zero constraint to the respective drive mode. For example, to disable AP mode, we set $x^{\rm ap}_t = 0$ for all $t$. In particular, we denote by \textsc{DMOP}$_{\rm cs}^{\rm ev}$ that has only EV and CS modes, without CE and AP modes, which can model series hybrid Chevrolet Volt.

\begin{figure*}[htp!]
\begin{minipage}[t]{0.275\textwidth}
\scriptsize
\noindent\rule{\textwidth}{1pt} 
\mbox{{\bf Algorithm 1.} ${\tt DMOP.DP}\big[G_0, B_0]$}
\noindent\rule{\textwidth}{0.4pt}
\begin{algorithmic}[1]
\State $F^\ast_T \leftarrow \infty$	
\For{each $\hat{B} \in [\underline{B}, \overline{B}]$}
\State $\big(F_T, (x_\tau)_{\tau=1}^{T} \big) \leftarrow {\tt DP}\big[\hat{B}, T, B_0 \big]$
\If{$F^\ast_T > F_T$}
\State $F^\ast_T \leftarrow F_T$
\State $(x^\ast_\tau)_{\tau=1}^{T} \leftarrow (x_\tau)_{\tau=1}^{T}$
\EndIf
\EndFor
\If{$F^\ast_T \le G_0$}
\State \Return $\big(F^\ast_T, (x^\ast_\tau)_{\tau=1}^{T} \big)$
\Else
\State \Return {\sf INFEASIBLE}
\EndIf  
\end{algorithmic}
\noindent\rule{\textwidth}{1pt}
\end{minipage}
\begin{minipage}[t]{0.325\textwidth}
\scriptsize
\noindent\rule{\textwidth}{1pt} 
\mbox{{\bf Algorithm 2.} ${\tt DP}\big[\hat{B}, t, B_0 \big]$}
\noindent\rule{\textwidth}{0.4pt}
\begin{algorithmic}[1]
\If{$t > 2$}

\State ${\sf cost}_t^{\min} \leftarrow \infty$

\For{each $B' \in [\underline{B}, \hat{B}]$}

\State $(Q_t, x_t) \leftarrow {\tt Solve}\big[\mbox{\rm\textsc{DMOP}}[\hat{B}, {B}', t] \big]$

\State $\big(F_{t\mbox{-}1}, (x_\tau)_{\tau=1}^{t\mbox{-}1} \big) \leftarrow {\tt DP}\big[B', t-1, B_0 \big]$

\If{${\sf cost}_t^{\min} > F(Q_t) + F_{t\mbox{-}1}$}
\State ${\sf cost}_t^{\min} \leftarrow F(Q_t) + F_{t\mbox{-}1}$
\State $F_t \leftarrow F(Q_t) + F_{t\mbox{-}1}$
\State $x \leftarrow \big((x_\tau)_{\tau=1}^{t\mbox{-}1}, x_t\big)$
\EndIf
\EndFor
\State \Return $\big(F_{t}, (x_\tau)_{\tau=1}^{t} \big)$
\Else
\State \Return ${\tt Solve}\big[\mbox{\rm\textsc{DMOP}}[\hat{B}, {B}_0, 1] \big]$
\EndIf  
\end{algorithmic}
\noindent\rule{\textwidth}{1pt}
\end{minipage}
\begin{minipage}[t]{0.4\textwidth}
\scriptsize
\noindent\rule{\textwidth}{1pt} 
\mbox{{\bf Algorithm 3.} ${\tt Online}\big[\theta^{\rm ap}, \theta^{\rm cs}, t, ( P_t, \eta^{\rm r}_t,$ $\eta^{\rm d}_t, \eta^{\rm e}_t, C_t, \beta_t ) \big]$}
\noindent\rule{\textwidth}{0.5pt}
\begin{algorithmic}[1]
\State $x^{\rm ev}_t \leftarrow 0, x^{\rm ce}_t \leftarrow 0, x^{\rm cs}_t \leftarrow 0, x^{\rm ap}_t \leftarrow 0$
\State \mbox{$r_t \leftarrow \min\{P^-_t, \frac{\overline{B} - B_{t\mbox{-}1}}{\eta^{\rm e}_t } \}$, $\tilde{s} \leftarrow \min\{\beta_t P^+_t, \frac{B_{t\mbox{-}1} - \underline{B}}{\eta^{\rm d}_t } \}$}
\State $\tilde{u} \leftarrow \min\{C_t, \frac{\overline{B} - B_{t\mbox{-}1} - \eta^{\rm r}_t r_t}{\eta^{\rm e}_t } \}$
\If{${\tt EV\_present}$ and $P^+_t \le \frac{B_{t\mbox{-}1} - \underline{B}}{\eta^{\rm d}_t}$}
\State $x^{\rm ev}_t \leftarrow 1$, $u_t \leftarrow 0$, $s_t \leftarrow P^+_t$ \Comment{{\em Use EV mode}}
\ElsIf{${\tt AP\_present}$ and $\big(\frac{F(P^+_t - \tilde{s})}{P^+_t - \tilde{s}} \le \theta^{\rm ap}$ or not (${\tt CS\_present}$ and ${\tt CE\_present})\big)$}
\State $x^{\rm ap}_t \leftarrow 1$, $u_t \leftarrow 0$, $s_t \leftarrow \tilde{s}$ \Comment{{\em Use AP mode}}
\ElsIf{${\tt CS\_present}$ and $\big(\frac{F(P^+_t + \tilde{u})}{P^+_t + \eta^{\rm e}_t \tilde{u}} \le \theta^{\rm cs}$ or not ${\tt CE\_present}\big)$}
\State $x^{\rm cs}_t \leftarrow 1$,  $u_t \leftarrow \tilde{u}$, $s_t \leftarrow 0$ \Comment{{\em Use CS mode}}
\ElsIf{${\tt CE\_present}$}
\State $x^{\rm ce}_t \leftarrow 1$,  $u_t \leftarrow 0$, $s_t \leftarrow 0$ \Comment{{\em Use CE mode}}
\EndIf
\State \Return $\big(x^{\rm ev}_t, x^{\rm ce}_t, x^{\rm cs}_t, x^{\rm ap}_t, r_t, s_t, u_t\big)$ 
\end{algorithmic}
\noindent\rule{\textwidth}{1pt}
\end{minipage} 
\end{figure*}

\section{Offline Solution} \label{sec:solution}

\subsection{Exact Solution by Dynamic Programming} \label{sec:dp}

This section provides offline solutions to solve \textsc{DMOP}.
Our exact solution is based on dynamic programming.

Consider a sub-problem $(\textsc{DMOP}[B_{t\mbox{-}1}, B_t, t])$ at time $t$, given the previous SoC $B_{t\mbox{-}1}$ and intended current SoC $B_t$:
\begin{align} \hspace{-10pt}
(\textsc{DMOP}[B_t, B_{t\mbox{-}1}, t]) & \qquad \displaystyle \min_{x_t}  \quad  F(Q_t) \notag\\
\mbox{subject to\ } &  \mbox{\ Cons.~\eqref{con:dmop-q}-\eqref{con:dmop-x0}} \notag
\end{align}
Note that given a fixed drive mode $(x_t)$, $\textsc{DMOP}[B_t, B_{t\mbox{-}1}, t]$ can be solved straightforwardly. Hence, $\textsc{DMOP}[B_t, B_{t\mbox{-}1}, t]$ can be solved by selecting the minimum-cost solution among the four drive modes. Let ${\tt Solve}\big[\mbox{\rm\textsc{DMOP}}[B_t, B_{t\mbox{-}1}, t] \big]$ be the minimum-cost solution, denoted by $(Q_t, x_t)$. If there is no feasible solution, ${\tt Solve}$ $\big[\mbox{\rm\textsc{DMOP}}[B_t, B_{t\mbox{-}1}, t] \big]$ returns infinite cost $Q_t$. Assuming the range $[\underline{B}, \overline{B}]$ is divided into a set of discrete levels, such that we can enumerate each possible level. A dynamic programming approach is presented in Algorithm ${\tt DMOP.DP}$. A similar approach can be applied to \textsc{DMOP} with absent drive modes (e.g., \textsc{DMOP}$_{\rm cs}^{\rm ev}$) by restricting each $\mbox{\rm\textsc{DMOP}}[B_t, B_{t\mbox{-}1}, t]$ to only the available drive modes.

\medskip

\begin{customthm}{1}\label{thm:t1}
Algorithm ${\tt DMOP.DP}$ provides an optimal solution for \textsc{DMOP} with pseudo-polynomial running time\footnote{The running time of a pseudo-polynomial-time algorithm depends polynomially on the size of input in unary representation, whereas the running time of a polynomial-time algorithm depends polynomially on the size of input in binary representation.}. The proof can be found in the Appendix.
\end{customthm}

\subsection{Approximation Solution by Convex Relaxation}  \label{sec:relax}

\textsc{DMOP} is a non-convex problem, even if we relax the integrality Cons.~\eqref{con:dmop-x0}, because of Cons.~\eqref{con:dmop-r}-\eqref{con:dmop-u}. Thus, we define a {\em convexified} problem (\textsc{cDMOP}) in the following way: 
\begin{enumerate}

\item We replace Cons.~\eqref{con:dmop-r}-\eqref{con:dmop-u} by simple upper bounds in Cons.~\eqref{con:cdmop-r}-\eqref{con:cdmop-u}. Cons.~\eqref{con:cdmop-bb} can ensure $(r_t, s_t, u_t)$ in \textsc{cDMOP} are not larger than those in \textsc{DMOP}. 

\item We add a constraint ($x^{\rm ev}_t P^+_t \le s_t$) to ensure the load to be entirely satisfied by electric motor in EV mode. 

\item We relax the decision variables $(x^{\rm ev}_t, x^{\rm ce}_t, x^{\rm cs}_t, x^{\rm ap}_t)$ to be fractions between 0 and 1 in Cons.~\eqref{con:cdmop-x}.

\end{enumerate}
By rounding the largest fractional variable among $(x^{\rm ev}_t, x^{\rm ce}_t,$ $x^{\rm cs}_t, x^{\rm ap}_t)$ up to one and other smaller variables down to zero (subject to the respective feasibility constrain of each mode), this provides a fast approximation solution to solve \textsc{DMOP}. 

\section{Online Solution}  \label{sec:online}

We present an online algorithm (Algorithm ${\tt Online}$) for $\textsc{DMOP}$ that does not require the future driving profile, but only the current vehicle state. Let the inputs of $\textsc{DMOP}$ be $(\sigma_t)_{t=1}^T = ( P_t, \eta^{\rm r}_t,$ $\eta^{\rm d}_t, \eta^{\rm e}_t, C_t, \beta_t )_{t=1}^T$. $\textsc{DMOP}$ can be solved optimally, when all inputs $\sigma$ are given in advance. However, $\sigma$ is revealed gradually over time in practice, which requires decision-making without future information.

An algorithm is called {\em online}, if the decision at the current time only depends on the information before or at the current time slot $t_{\rm now}$ (i.e., $( \sigma_t )_{t\le t_{\rm now}}$).
Given input $\sigma$, let ${\sf Cost}({\tt Alg}[\sigma])$ be the cost of algorithm ${\tt Alg}$, and ${\sf Opt}(\sigma)$ be the cost of an offline optimal solution (that may rely on an oracle to obtain all future inputs). In competitive analysis for online algorithms \cite{allan1998occa}, {\em competitive ratio} is a metric defined by the {\em worst-case} ratio between the cost of the online algorithm ${\tt Alg}$ and that of an offline optimal solution, namely,
\begin{equation}
{\tt CR}({\tt Alg}) \triangleq \max_{\sigma} \frac{{\sf Cost}({\tt Alg}[\sigma])}{{\tt Opt}(\sigma)}
\end{equation}

Algorithm ${\tt Online}$ selects the drive mode following priority: EV $\to$ AP $\to$ CS $\to$ CE, based on the respective conditions. Define {\em normalized cost} by the fuel cost over the amount of used fuel (i.e., $\frac{F(P^+_t + u_t)}{P^+_t + \eta^{\rm e}_t u_t}$ in CS mode and $\frac{F(P^+_t - s_t)}{P^+_t - s_t}$ in AP mode). First, EV mode is selected, if it is present and there is sufficient SoC. Else, AP mode is selected, if it is present and the normalized cost is less than threshold $\theta^{\rm ap}$ or CS and CE modes are not present. Else, CS mode is selected, if it is present and the normalized cost is less than $\theta^{\rm cs}$ or CE mode is not present. Otherwise, CE mode is selected. The basic idea of ${\tt Online}$ is to make conservative decision at each time step, such that the incurred energy consumption is within a certain bounded range from the offline optimal solution.

Define the per-unit cost by $f(Q) \triangleq \tfrac{F(Q)}{Q}$. Note that $F(\cdot)$ is strictly increasing convex and $f(Q)$ is an increasing function. Suppose $f_{\min} \le f(Q) \le f_{\max}$ for some constants $f_{\min} \ge f(0)$ and $f_{\max}\le f(\overline{G})$. We assume $f_{\min}, f_{\max}$ can be estimated in advance for a specific trip\footnote{First, set $f_{\min} = f_{\max} = f(Q_1)$. Then $f_{\min}, f_{\max}$ are updated to be the maximum and minimum $f(Q_t)$ observed so far at each time $t$. If $T$ is relatively large, the estimated $f_{\min}, f_{\max}$ will converge to the true values.}. Let
$
\eta^{\rm d}_{\min} \triangleq \min_{t} \eta^{\rm d}_t, \
\eta^{\rm e}_{\min} \triangleq \min_{t} \eta^{\rm e}_t, \
\eta^{\rm e}_{\max} \triangleq \max_{t} \eta^{\rm e}_t.
$
Next, we determine proper $\theta^{\rm ap}$ and $\theta^{\rm cs}$ with a good competitive ratio.

\medskip

\begin{customthm}{2} \label{thm:online}
We consider the initial SoC $B_0 = \underline{B}$ and we require the final SoC to be $B_{T+1} = \overline{B}$.
Assuming $P^-_t = 0$ for all $t$, let the thresholds in Algorithm ${\tt Online}$ be
\begin{equation}
\theta^{\rm cs} = \textstyle \sqrt{\frac{f_{\max}f_{\min}}{\kappa \eta^{\rm d}_{\min} \eta^{\rm e}_{\max}}}
\quad \mbox{\ and\ } \quad \theta^{\rm ap} = \theta^{\rm cs} \eta^{\rm d}_{\min} \eta^{\rm e}_{\min},
\end{equation}
where $\kappa \triangleq \max\{1,\frac{1}{\eta^{\rm e}_{\max}\eta^{\rm d}_{\min}} \}$,
then the competitive ratio of ${\tt Online}$ for solving $\textsc{DMOP}$ is  
\begin{equation}
{\tt CR}({\tt Online}) = \textstyle \sqrt{\frac{\kappa f_{\max} \eta^{\rm e}_{\max}}{f_{\min} \eta^{\rm d}_{\min}}} \frac{1}{\eta^{\rm e}_{\min}}.
\end{equation}
\end{customthm}

\section{Integrated Path Planning}

We consider an integrated optimization problem with both path planning and drive mode decisions for the PHEV:
\begin{itemize}
\item There are multiple paths between source and destination.
\item There are possible intermediate nodes in each path to provide fuel refilling or battery charging.
\end{itemize}
A road network is represented by a connected directed graph ${\cal G} = ({\cal N}, {\cal E})$ that connects from the source ${\sf v_s}$ to the destination ${\sf v_d}$. For each edge $e=({\sf u},{\sf  v}) \in {\cal E}$, ${\sf u}$ may be a stop, such that the PHEV can receive refilling of fuel at price $g_{\sf u}$ per unit, or battery charging at most $E_{\sf u}$ unit at price $h_{\sf u}$ per unit.
Let ${\cal P}$ be the set of paths connecting ${\sf v}_{\sf s}$ and ${\sf v_d}$. We label the edges in each $P \in {\cal P}$ by $(e_1, e_2, ..., e_{n(P)})$ and write $e_i=({\sf u}_i,{\sf v}_i)$.
Let $T(e)$ be the number of time slots required for traveling $e$. Let $G_{e,t}$ be the initial fuel tank level at time $t$, when traveling $e$. Let the fuel tank capacity be $\overline{G}$.

The path planning problem with drive mode decisions is formulated as an integer programming problem (\textsc{PPDM}).

\begin{figure}[htp!]
\hspace{-10pt} 
\begin{align}
&(\textsc{PPDM}) \ \displaystyle \min_{P \in {\cal P}, (x_{e,t})_{t=1, e\in P}^{T(e)}}  \quad \sum_{i=1}^{n(P)} g_{{\sf u}_i} (G_{e_i,0} - G_{e_{i\mbox{-}1},T({e_{i\mbox{-}1}})}) \notag\\
& \qquad \qquad \qquad \qquad \qquad \qquad \qquad +  h_{{\sf u}_i} (B_{e_i,0}-B_{e_{i\mbox{-}1},T({e_{i\mbox{-}1}})} ) \notag\\
& \mbox{subject to for all\ }  e \in {\cal E}, t \in [1, T({e})] \notag
\end{align}
\hspace{-10pt}
\begin{align}
 & \quad G_{e,0} \le \overline{G}, \label{con:pdmp-f00}\\
 & \quad G_{e,t} = G_{e,t\mbox{-}1} - F(Q_{e,t}), \label{con:pdmp-f} \\
 & \quad G_{e,t} \ge 0, \label{con:pdmp-f0}\\
 & \quad Q_{e,t} = (1 - x^{\rm ev}_{e,t})P^+_{e,t} + x^{\rm cs}_{e,t} u_{e,t} - x^{\rm ap}_{e,t} s_{e,t}, \label{con:pdmp-q}\\
 & \quad B_{e,t} = B_{e,t\mbox{-}1} + \eta^{\rm r}_{e,t} r_{e,t}+ \eta^{\rm e}_{e,t} u_{e,t} - \eta^{\rm d}_{e,t} s_{e,t}, \label{con:pdmp-b} \\
 & \quad \underline{B} \le  B_{e,t} \le \overline{B}, \label{con:pdmp-bb}\\
 & \quad B_{e',T(e')}- B_{e,0} \le E_{{\sf u}}, \mbox{if\ } ({\sf u}, {\sf v}) = e \mbox{\ and\ } ({\sf v}', {\sf u}) = e' \label{con:pdmp-b0} \\
 & \quad r_{e,t} = \min\{P^-_{e,t}, \textstyle \frac{\overline{B} - B_{e,t\mbox{-}1}}{\eta^{\rm r}_{e,t} } \} \label{con:pdmp-r}\\
 & \  s_{e,t} = \min\{x^{\rm ev}_{e,t} P^+_{e,t} + x^{\rm ap}_{e,t} \beta_{e,t} P^+_{e,t}, \textstyle \frac{B_{e,t\mbox{-}1} - \underline{B}}{\eta^{\rm d}_{e,t} } \} \label{con:pdmp-s}\\
 & \quad u_{e,t} = \min\{x^{\rm cs}_{e,t} C_{e,t}, \textstyle \frac{\overline{B} - B_{e,t\mbox{-}1} - \eta^{\rm r}_{e,t} r_{e,t}}{\eta^{\rm e}_{e,t} } \} \label{con:pdmp-u}\\
 & \quad x^{\rm ev}_{e,t} + x^{\rm ce}_{e,t} + x^{\rm cs}_{e,t} + x^{\rm ap}_{e,t} = 1, \label{con:pdmp-x}\\
 & \quad x^{\rm ev}_{e,t}, \ x^{\rm ce}_{e,t}, \ x^{\rm cs}_{e,t}, \ x^{\rm ap}_{e,t} \in \{0, 1\} \label{con:pdmp-x0}
\end{align}  
\end{figure}

In \textsc{PPDM}, we find the optimal path $P \in {\cal P}$ from ${\sf v_s}$ to ${\sf v_d}$ and the corresponding drive modes $(x_{e,t})_{t=1, e\in P}^{T(e)}$, given the initial fuel level ($G_{e_0,T(e_0)} = G_0$) and SoC ($B_{e_0,T(e_0)} = B_0$).

\subsection{Exact Solution by Dynamic Programming} \label{sec:dp2}

\subsubsection{Uniform Cost Case}
First, we consider the {\it uniform} case with identical fuel price ($g_{\sf u}=1$ and $h_{\sf u}=0$ for all ${\sf u}$). For a path $P=(e_1,...,e_{n(P)})\in {\cal P}$, by Eqn.~\eqref{con:pdmp-f}, we obtain 
\begin{equation}\label{e1}
G_{e_i,0}-G_{e_i,T({e_i})}=\sum_{t=1}^{T({e_i})}F(Q_{e_i},t), \mbox{\ for\ }i=2,...,n(P) 
\end{equation}
By Eqn.~\eqref{con:pdmp-f00} and Eqn.~\eqref{con:pdmp-f0}, we obtain 
\begin{equation}\label{e2}
\sum_{t=1}^{T({e_i})}F(Q_{e_i},t)\le \overline G 
\end{equation}
Thus, we can rewrite \textsc{PPDM} as follows:  
\begin{align}
(\textsc{uPPDM}) & \ \displaystyle \min_{(x_{e,t})_{t=1, e\in P}^{T(e)}, P \in {\cal P}}  \quad \sum_{i=1}^{n(P)} \sum_{t=1}^{T({e_i})} F(Q_{e_i,t})\notag\\
\mbox{subject} & \mbox{\ to Eqns.~\eqref{e2} and \eqref{con:pdmp-q}-\eqref{con:pdmp-x0} \ }  \notag  
\end{align}

\begin{figure}[htb!]
	\centering  
    \includegraphics[scale = 0.6]{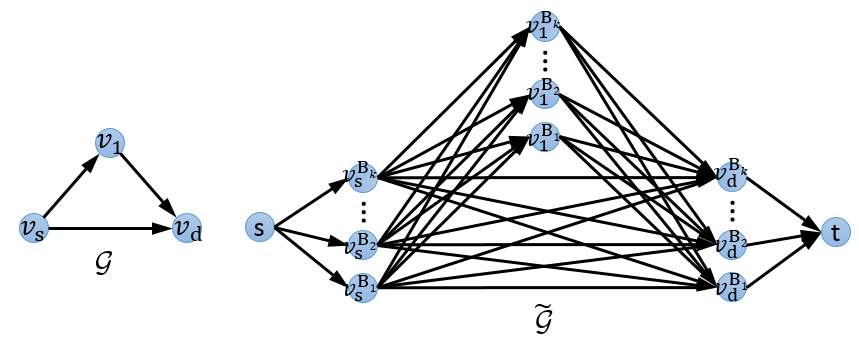} 
    \caption{An illustration of ${\mathcal G}$ and $\widetilde{\mathcal G}$, where $[\underline B,\overline B]$ is a set of discrete levels $\{B_1, ..., B_k \}$.}\label{fig:path-ex} 
\end{figure}

To solve \textsc{uPPDM}, we construct a weighted directed graph $\widetilde{\mathcal G}=(\widetilde{\mathcal N},\widetilde{\mathcal E},w)$ as follows (see Fig.~\ref{fig:path-ex}). Let $z^\ast_{e}(B,B')$ be the value of the optimal solution of \textsc{DMOP} for edge $e\in\cal{E}$, when the $B_{e,0}=B$ and $B_{e,T(e)}=B'$, which can be obtained by dynamic programming as explained in Sec.~\ref{sec:dp}.
For every node ${\sf v}\in \cal{N}$ and every discrete level $B$ in the range $[\underline B,\overline B]$, we create a node ${\sf v}^B\in\widetilde{\mathcal N}$. If $e=({\sf u},{\sf v})\in\cal {E}$ then we have an edge $({\sf u}^B,{\sf v}^{B''})\in\widetilde{\mathcal E}$ with weight $w({\sf u}^B,{\sf v}^{B''})=z^\ast_{e}(B,B')$, for every\footnote{Note that, since battery charging at each node is free, it suffices to construct only one edge corresponding to $B''=\min\{B'+E_{\sf u},\overline B\}$; however, defining the graph in this general form allows to extend the dynamic program for the case when there is a cost for charging each unit of battery at node ${\sf u}$.} discrete levels $B,B',B''$ in $[\underline B,\overline B]$ such that 
\begin{equation}\label{eq:B-range}
z^\ast_{e}(B,B')\le \overline G \mbox{ \ and \ } B'\le B''\le B'+E_{\sf u}
\end{equation}
In addition, we create a source node ${\sf s}\in\widetilde{\mathcal N}$ with edges $({\sf s},{\sf v}_{\sf s}^{B})$, for each $B$ in the range $[B_0,\min\{B_0+E_{\sf v_s},\overline{B}\}]$, having weight $\overline{G}-G_0$ and cost $g_{\sf s}=0$; and a destination node ${\sf t}\in\widetilde{\mathcal N}$ with edges $({\sf v}_{\sf d}^B,{\sf t})$ having weight $0$, for all $B$ in the range $[\underline B,\overline B]$.

Then the optimal solution to \textsc{uPPDM} can be obtained by finding an $({\sf s,t})$-shortest path in the graph $\widetilde{\mathcal G}$, with the (non-negative) weights $w(\cdot,\cdot)$ interpreted as distances.

\subsubsection{General Case}

Next, we consider the more general case when the fuel cost per unit $g_{\sf u}$ may not be equal at all nodes ${\sf u}\in\mathcal{N}$, and with the additional restriction that the PHEV can make at most $\Delta$ stops between ${\sf v_s}$ and ${\sf v_d}$ to do a fuel refill. We assume that battery charging (at the stop) is allowed only when the vehicle stops for fuel refill\footnote{This assumption can be removed, if $h_{\sf u} = 0$ at all nodes.}.also, we assume that battery charging at node ${\sf u}$ costs $h_{\sf u}$ per unit, and the objective is to minimize the combined fuel cost and battery charging cost.
The basic idea is to adopt the dynamic program for the so-called {\it Gas Station Problem} in \cite{samir2011fillgas}, and apply it to the graph $\widetilde{\mathcal G}$ constructed above.
We define the graph $\mathcal{G}_0$ as the subgraph of $\widetilde{\mathcal G}$ such that $E_e=0$ for all $e$. (That is, in \eqref{eq:B-range}, $B''$ takes only one value, namely $B'$, which corresponds to the case when no battery charging is allowed at node ${\sf u}$.)

As we have discretized the battery level in $[\underline B, \overline B]$, we may also discretize the gas level in $[0,\overline G]$. However, following \cite{samir2011fillgas}, we can already define a discrete set ${\mathscr G}({\sf v}^B)$, as the set of fuel levels that are sufficient to consider at node ${\sf v}^B\in\widetilde{\mathcal N}$:
\begin{align}\label{eq:e11} \hspace{-10pt}
{\mathscr G}({\sf v}^B)&\triangleq \Big\{\overline G-w({\sf u}^B,{\sf v}^{B'})~|~{\sf v^{B'}}\in \widetilde{\mathcal N},~g_{\sf u}<g_{\sf v}\nonumber\\
&\qquad\text{ and } w({\sf u}^B,{\sf v}^{B'})<\overline G
\Big\}
\cup\{0\} 
\end{align}
See Lemma~\ref{lem:l1} in the Appendix for a justification. Here, $w({\sf u}^{B},{\sf v}^{B'})$ is the shortest distance (with respect to the weights $w$) between (not necessarily adjacent nodes) ${\sf u}^{B}$ and ${\sf v}^{B'}$ in the graph $\mathcal{G}_0$ (note that we use $\mathcal{G}_0$ as it is assumed that the PHEV does not make any stop between ${\sf u}^{B}$ and ${\sf v}^{B'}$).

\begin{figure*}[!t]
\begin{align}
C[{\sf u}^B,1,g]
&=\left\{\begin{array}{cl}
\displaystyle \min_{B\le B'\le\min\{B+E_{\sf u},\overline B\}}\Big((w({\sf u}^{B'},{\sf t})-g)g_{\sf u}+h_{\sf u}(B'-B)\Big), & \text{ if } g\le w({\sf u}^{B'},{\sf t})\le \overline {G}\\
\infty, & \text{ otherwise} \\
\end{array}\right. \label{recurrence1} \\
C[{\sf u}^B,q,g]
&= \min_{\stackrel{B\le B'\le\min\{B+E_{\sf u},\overline B\},{\sf v}^{B''}:}{w({\sf u}^{B'},{\sf v^{B''}})\le \overline{G}}}
\left\{\begin{array}{@{}l@{}l@{}}
C[{\sf v}^{B''},q - 1,0]+(w({\sf u}^{B'},{\sf v}^{B''})-g)g_{\sf u}+h_{\sf u}(B' - B),
 \text{ if }g_{\sf v}\le g_{\sf u}\text{ and }g\le w({\sf u}^{B'},{\sf v}^{B''})\\ \\
C\Big[{\sf v}^{B''},q - 1,\overline G  -  w({\sf u}^{B'},{\sf v}^{B''})\Big]+(\overline G - g)g_{\sf u}+h_{\sf u}(B' - B),
\text{ if }g_{\sf v}> g_{\sf u}
\end{array}\right. \label{recurrence2}
\end{align} 
\end{figure*}

Let $C[{\sf u}^B,q,g]$ be the minimum cost of going (in the graph $\widetilde{\mathcal G}$) from ${\sf u}_B$ to ${\sf t}$ using $q$ stops (including ${\sf u}^{B}$), when the fuel level at ${\sf u}^B$ is $g$. Then we can write the recurrence Eqns.~\eqref{recurrence1}-\eqref{recurrence2} for $C[{\sf u}^B,q,g]$ for any $g\in {\mathscr G}({\sf u}^{B})$ and $2\le q\le\Delta$ (see the proof of Theorem~\ref{thm:PPDM} for an explanation). A dynamic programming based algorithm is described in ${\tt PPDM.DP}$.

\begin{figure}[htp!]
\scriptsize
\noindent\rule{0.47\textwidth}{1pt} 
\mbox{{\bf Algorithm 4.} ${\tt PPDM.DP}\big[G_0, B_0]$} \\
\noindent\rule{0.47\textwidth}{0.4pt}
	\begin{algorithmic}[1]
	\State Construct graph $\widetilde{\mathcal G}=(\widetilde{\mathcal N},\widetilde{\mathcal E},w)$ and extract subgraph $\mathcal{G}_0$
	\State Let $\mathcal{B}$ be the discretized range between $\underline{B}$ and $\overline{B}$
	\State Find all-pairs shortest-distances $\{w(u^B,v^{B'})\}_{u^B,v^{B'}}$ in $\mathcal{G}_0$
    \For{each ${B}\in\mathcal{B}$, ${\sf v}\in \mathcal{N}$}
    \State Let ${\mathscr G}({\sf v}^{B})$ be as given by Eqn.~\eqref{eq:e11}
    \EndFor
	\For{each ${B}\in\mathcal{B}$, ${\sf u}\in \mathcal{N}$, $q\in \{1,...,\Delta\}$ and $g\in {\mathscr G}({\sf u}^{B})$}
	  \State $C[u^B,q,g]\leftarrow\infty$
	\EndFor
	\LeftComment{ {\em Compute Eqn.~\eqref{recurrence1}}}
	\For{each ${B}\in\mathcal{B}$, ${\sf u}\in \mathcal{N}$, and $g\in {\mathscr G}({\sf u}^{B})$}
	  \For{$B'\in\mathcal{B}$ such that $B\le B'\le\min\{B+E_{\sf u},\overline B\}$}
	  \If{$g\le w({\sf u}^{B'},{\sf t})\le \overline {G}$ and \\
	  \qquad \qquad  $(w({\sf u}^{B'},{\sf t})-g)g_{\sf u}+h_{\sf u}(B'-B)<C[u^B,1,g]$}
	
		\State \qquad \quad $C[u^B,1,g]\leftarrow(w({\sf u}^{B'},{\sf t})-g)g_{\sf u}+h_{\sf u}(B'-B)$
	  \EndIf
	\EndFor
	\EndFor
	\LeftComment{ {\em Compute Eqn.~\eqref{recurrence2}}}
	\For{each ${B}\in\mathcal{B}$, ${\sf u}\in \mathcal{N}$, $q\in\{1,...,\Delta\}$, and $g\in {\mathscr G}({\sf u}^{B})$}
	  \For{$B'\in\mathcal{B}$ such that $B\le B'\le\min\{B+E_{\sf u},\overline B\}$}
		\For{$B''\in\mathcal{B}$ and ${\sf v}\in \mathcal{N}$ such that $w({\sf u}^{B'},{\sf v^{B''}})\le \overline{G}$}
		   \If{$g_{\sf v}\le g_{\sf u}$, $g\le w({\sf u}^{B'},v^{B''})$ and $C[{\sf v}^{B''},q - 1,0]$ \\
		   \quad \qquad \qquad $+(w({\sf u}^{B'},{\sf v}^{B''}) - g)g_{\sf u}+h_{\sf u}(B' - B)<C[u^B,q,g]$}
			
			  \State $C[u^B,q,g]\leftarrow C[{\sf v}^{B''},q - 1,0]+(w({\sf u}^{B'},{\sf v}^{B''}) - g)g_{\sf u}$
			  \Statex \qquad \qquad \qquad \qquad \qquad \qquad \qquad $+ h_{\sf u}(B' - B)$
		   \Else
			  \If{$g_{\sf v}> g_{\sf u}$ and $C\Big[{\sf v}^{B''},q - 1,\overline G  -  w({\sf u}^{B'},{\sf v}^{B''})\Big]$ \\
			 \qquad \qquad \qquad \qquad \qquad $+(\overline G - g)g_{\sf u}+h_{\sf u}(B' - B)<C[u^B,q,g]$}
			
			  \State $C[u^B,q,g]\leftarrow C\Big[{\sf v}^{B''},q-1,\overline G  -  w({\sf u}^{B'},{\sf v}^{B''})\Big]$
			  \Statex \qquad \qquad \qquad  \qquad \qquad \qquad \qquad \qquad $+(\overline G - g)g_{\sf u}+h_{\sf u}(B' - B)$
			\EndIf
		   \EndIf
		\EndFor
	 \EndFor
   \EndFor\\  	
   \Return $\min_{1\le q\le \Delta}C[{\sf s},q,0]$
\end{algorithmic}
\noindent\rule{0.47\textwidth}{1pt}
\end{figure}

\begin{customthm}{3}\label{thm:PPDM}
Algorithm ${\tt PPDM.DP}$ computes an optimal solution for \textsc{PPDM} with pseudo-polynomial running time.
\end{customthm}

\subsection{Approximation Solution by Convex Relaxation} \label{sec:relax2}

We associate a variable $y_e\in\{0,1\}$ to each edge $e\in{\cal E}$ indicating whether or not the edge $y_e$ is selected in the optimal path. In addition to Cons.~\eqref{con:pdmp-f00}-\eqref{con:pdmp-x0}, we add the flow conservation constraints involving the variables $y_e$. \textsc{PPDM} can be formulated by an alternate formulation in \textsc{PPDM-IP}. 

\begin{figure}[htp!]
(\textsc{PPDM-IP}) 
\hspace{-10pt} 
\begin{align} 
&\displaystyle \min_{(x_{e,t},y_e)_{t=1, e\in {\cal E}}^{T(e)}}  \quad \sum_{{\sf u}\in {\cal N}} g_{\sf u}\bigg(\sum_{e=({\sf u},{\sf  v})}y_e G_{e,0} -\sum_{e=({\sf v},{\sf  u})}y_e G_{e,T(e)} \bigg) \notag\\
& \qquad \qquad \qquad  +  h_{{\sf u}}\bigg(\sum_{e=({\sf u},{\sf  v})}y_e B_{e,0} -\sum_{e=({\sf v},{\sf  u})}y_e B_{e,T(e)} \bigg)  \notag\\
& \mbox{subject to Eqns.~\eqref{e2}, \eqref{con:pdmp-q}-\eqref{con:pdmp-x0} and for all\ } e \in {\cal E}, t \in [1, T({e})] \notag\\
& \quad \sum_{e=({\sf u},{\sf  v})}y_e- \sum_{e=({\sf v},{\sf  u})}y_e=\left\{
\begin{array}{ll}
1 &\text{ if }{\sf u}={\sf s}\\
-1 &\text{ if }{\sf u}={\sf t}\\
0 &\text{ otherwise}
\end{array} 
\right. \label{con:ip-pdmp-cons}\\ 
& \quad \sum_{e=({\sf u},{\sf  v})}y_e B_{e,0} -\sum_{e=({\sf v},{\sf  u})}y_e B_{e,T(e)} \le E_{{\sf u}} \label{con:ip-pdmp-b0} \\
& \quad y_e\in\{0,1\} \text{ for all } e\in{\cal E}\label{con:ip-pdmp-y0}
\end{align} 
\end{figure}

Similar to \textsc{cDMOP}, \textsc{PPDM-IP} can be convexified as \textsc{cPPDM}, by relaxing the integrality constraint Cons.~\eqref{con:pdmp-x0}, relaxing the equality in Cons.~\eqref{con:pdmp-f}, relaxing Cons.~\eqref{con:pdmp-r}-\eqref{con:pdmp-u}, and linearizing the quadratic terms by substituting $\lambda_e^0=y_eG_{e,0}$, $\lambda_e=y_eG_{e,T(e)}$, $\mu_e^0=y_eB_{e,0}$, $\mu_e=y_eB_{e,T(e)}$. 

\begin{figure}[htp!]
(\textsc{cPPDM}) 
 \hspace{-10pt} 
 \begin{align} 
 &\displaystyle \min_{(x_{e,t},y_e,\lambda_e^0,\lambda_e,\mu_e^0,\lambda_e)_{t=1, e\in {\cal E}}^{T(e)}}  \quad \sum_{{\sf u}\in {\cal N}} g_{\sf u}\bigg(\sum_{e=({\sf u},{\sf  v})}\lambda_e^0-\sum_{e=({\sf v},{\sf  u})}\lambda_e \bigg) \notag\\
 & \qquad \qquad \qquad  +  h_{{\sf u}}\bigg(\sum_{e=({\sf u},{\sf  v})}\mu_e^0 -\sum_{e=({\sf v},{\sf  u})}\mu_e \bigg)  \notag\\
 & \mbox{subject to Eqns.~\eqref{con:pdmp-q}-\eqref{con:pdmp-bb} and for all\ } e \in {\cal E},{\sf u}\in{\cal N}, t \in [1, T({e})] \notag\\
 & \ \quad G_{e,t} \le G_{e,t\mbox{-}1} - F(\mu_{e,t}), \label{con:cpdmp-f} \\
 & \ \quad \sum_{e=({\sf u},{\sf  v})}\mu_e^0 -\sum_{e=({\sf v},{\sf  u})}\mu_e \le E_{{\sf u}} \label{con:cpdmp-b0} \\
 & \ \quad r_{e,t} \le P^-_{e,t}, \label{con:cpdmp-r}\\
 & \ \quad  x^{\rm ev}_{e,t} P^+_{e,t}\le s_{e,t} \le x^{\rm ev}_{e,t} P^+_{e,t} + x^{\rm ap}_{e,t} \beta_{e,t} P^+_{e,t}, \label{con:cpdmp-s}\\
 & \ \quad u_{e,t} \le x^{\rm cs}_{e,t} C_{e,t},  \label{con:cpdmp-u}\\
 & \ \quad x^{\rm ev}_{e,t} + x^{\rm ce}_{e,t} + x^{\rm cs}_{e,t} + x^{\rm ap}_{e,t} = 1, \label{con:cpdmp-x}\\ 
 & \ \quad x^{\rm ev}_{e,t}, \ x^{\rm ce}_{e,t}, \ x^{\rm cs}_{e,t}, \ x^{\rm ap}_{e,t} \in [0, 1] \label{con:cpdmp-x0}\\  
 &\ \quad 0\le \lambda_e^0\le \overline{G}y_e\label{con:cdmp-quad-G0-1}\\
&\ \quad G_{e,0}-\overline{G}(1-y_e)\le \lambda_e^0\le G_{e,0}\label{con:cdmp-quad-G0-2}\\
&\ \quad 0\le \lambda_e\le \overline{G}y_e\label{con:cdmp-quad-GT-1}\\
&\ \quad G_{e,T(e)}-\overline{G}(1-y_e)\le \lambda_e^0\le G_{e,T(e)}\label{con:cdmp-quad-GT2}\\
&\ \quad \underline{B}y_e\le \mu_e^0\le \overline{B}y_e\label{con:cdmp-quad-B0-1}\\
&\ \quad B_{e,0}-\overline{B}(1-y_e)\le \mu_e^0\le B_{e,0}-\underline{B}(1-y_e)\label{con:cdmp-quad-B0-2}\\
&\ \quad \underline{B}y_e\le \mu_e\le \overline{B}y_e\label{con:cdmp-quad-BT-1}\\
&\ \quad B_{e,T(e)}-\overline{B}(1-y_e)\le \mu_e\le B_{e,T(e)}-\underline{B}(1-y_e)\label{con:cdmp-quad-BT-2}\\
 & \ \quad y_e\in[0,1] \label{con:cpdmp-y0}
 \end{align} 
\end{figure}
 
\begin{figure*}[!t]
\subfigure[Estimated ${\eta^{\rm r, d}_t P_t}$ vs. $P_t^{\sf{B}}$.]{\label{fig:modelEM} \includegraphics[width=0.36\textwidth]{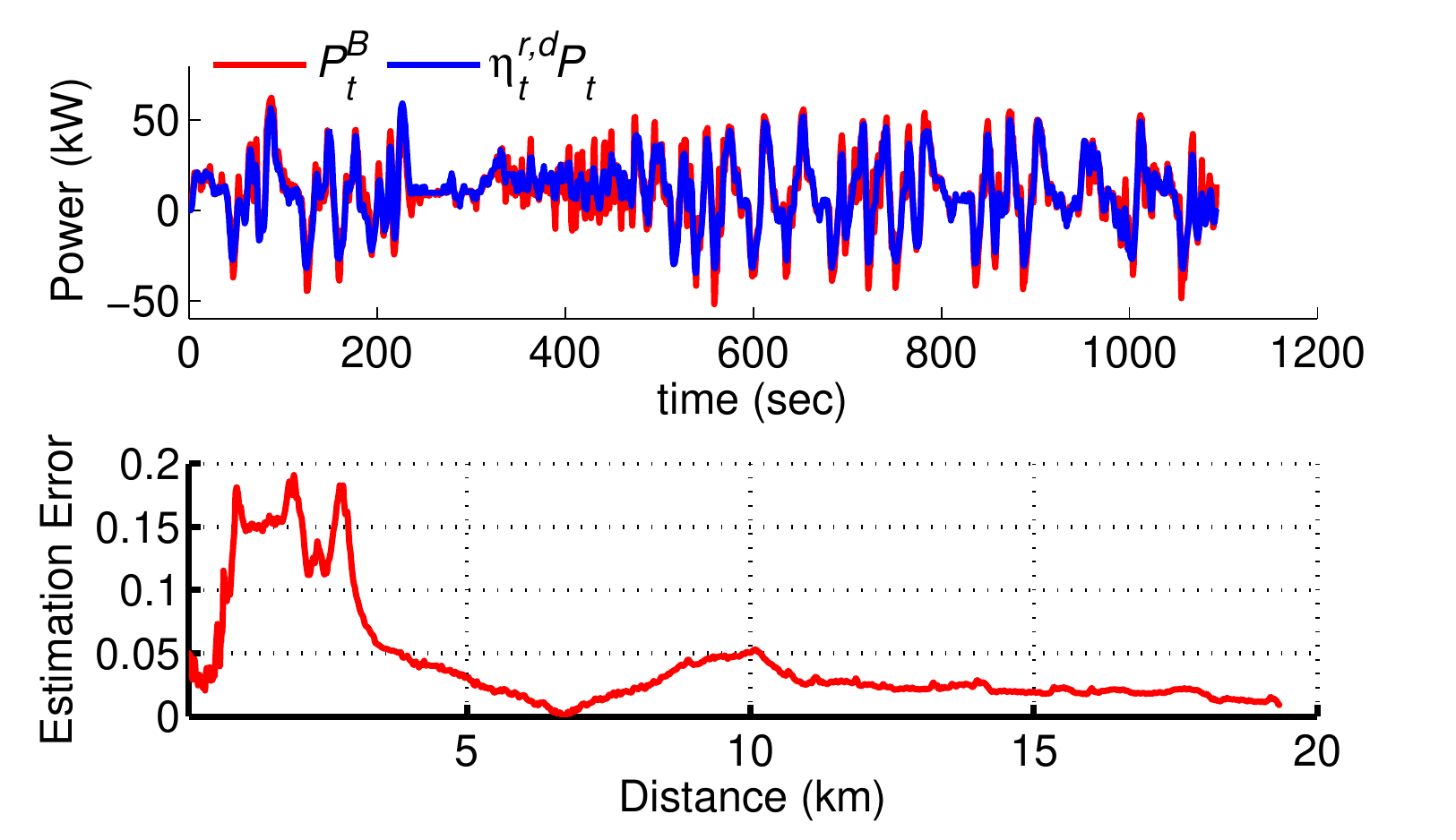}}\hspace{-2em}
\subfigure[Estimated ${\eta^{\rm e}_t} u_t$ vs. $u_t^{\sf{B}}$.]{\label{fig:modelGe} \includegraphics[width=0.36\textwidth]{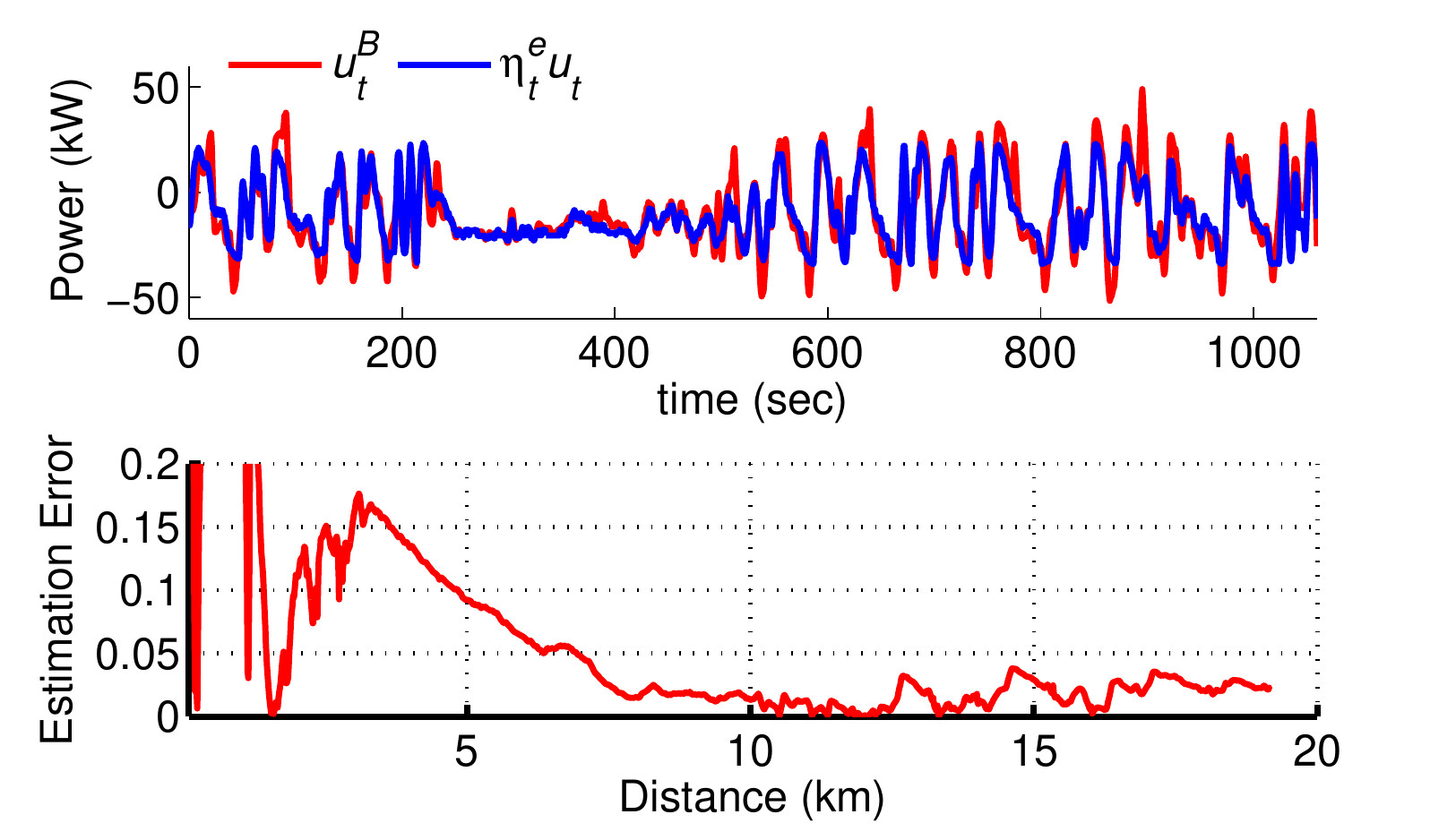}}\hspace{-2em}
\subfigure[Estimated $\hat{F}(\cdot)$ vs. ${F}(\cdot)$.]{\label{fig:modelEng} \includegraphics[width=0.36\textwidth]{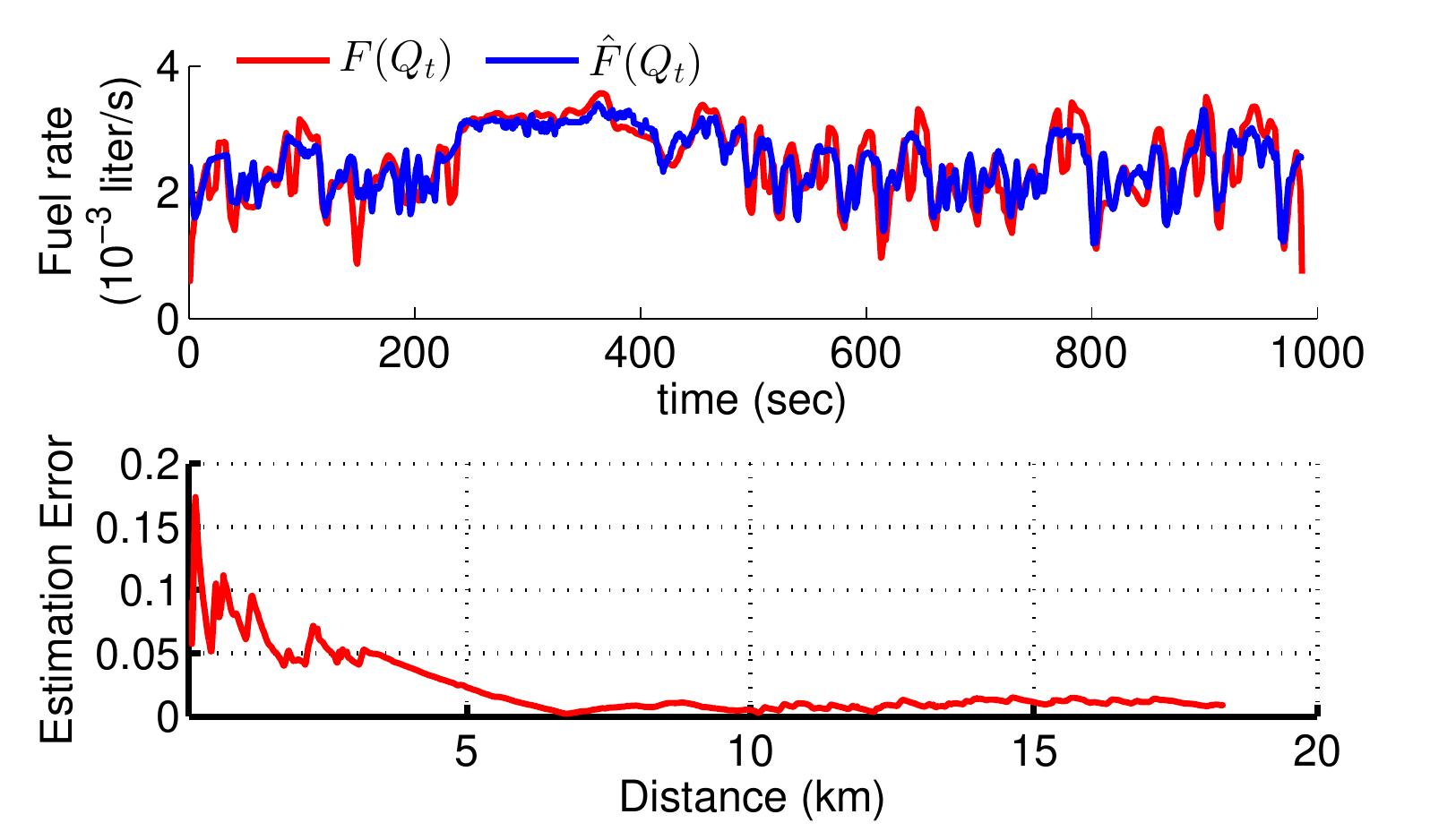}} 
\caption{Comparisons of the estimated values using regression models with the actual measurements.}\label{fig:modelfit}
\end{figure*}

\section{Evaluation} \label{sec:eval}
We evaluate our system empirically on a Chevrolet Volt (Model 2013). Validated by real-world data measured on Volt, we observe that our system can enable significant fuel savings.

\subsection{Vehicle Model Validation and Calibration}

We first estimate the efficiency coefficients ${\eta^{\rm r}_t}, {\eta^{\rm d}_t}, {\eta^{\rm e}_t}$ based on the driving profile. We utilize On-board Diagnostics (OBD) dongle to collect data from Chevrolet Volt (e.g., battery voltage, battery current, motor voltage, motor current, fuel rate and vehicle speed), which are used to calibrate the regression models that compute ${\eta^{\rm r}_t}, {\eta^{\rm d}_t}, {\eta^{\rm e}_t}$, given driving profile as input. Readers can refer to \cite{cmtseng2016EVextract} for the details of extraction OBD data. Our testing environment is relatively flat, and hence, we assume the gradient of road $\alpha_t =0$.
Table~\ref{tab:paramdescript} presents the settings of parameters for Chevrolet Volt in the model.

Let the measured power of battery be $P_t^{\sf{B}}$, which is related to the load of drivetrain by the following equations: 
\begin{equation}
P_t^{\sf{B}+}={P_t^+}{\eta^{\rm d}_t}, \quad P_t^{\sf{B}-}={P_t^-}{\eta^{\rm r}_t}
\end{equation}
We estimate ${\eta^{\rm r}_t}, {\eta^{\rm d}_t}$ by the following regression model: 
\begin{equation}
\eta_t^{r,d} = \lambda_1 v_t^2 + \lambda_2 v_t+\lambda_3 a^{+2}_2+\lambda_4 a^+_t \lambda_5 a^{-2}_t + \lambda_6 a^{-}_t + \lambda_7 \label{eqn:etaEVmodel}
\end{equation}
where $a^{+}_t \triangleq \max\{v_t - v_{t\mbox{-}1}, 0\}$ and $a^{-}_t \triangleq \max\{v_{t\mbox{-}1} - v_t, 0\}$. Note that if $a^{+}_t>0$, then $a^{-}_t=0$.

Let the measured power from combustion engine to charge the battery be $u_t^{\sf{B}}=\eta_t^e u_t$, where ${\eta^{\rm e}_t}$ is estimated by: 
\begin{equation}
\eta_t^e = \mu_1 v_t^2 + \mu v_t+\mu_3 a^{+2}_2+\mu_4 a^+_t \mu_5 a^{-2}_t + \mu_6 a^{-}_t + \mu_7  \label{eqn:pgenmodel}
\end{equation}

We also estimate fuel consumption function $F(\cdot)$ by: 
\begin{equation}
\hat{F}(Q_t)=\gamma_1Q_t^2+\gamma_2Q_t+\gamma_3 \label{eqn:FCmodel}
\end{equation}

\begin{figure*}[htb!] 
\subfigure[Case study of series hybrid on a route.]{\label{fig:opt1}\includegraphics[width=0.27\textwidth]{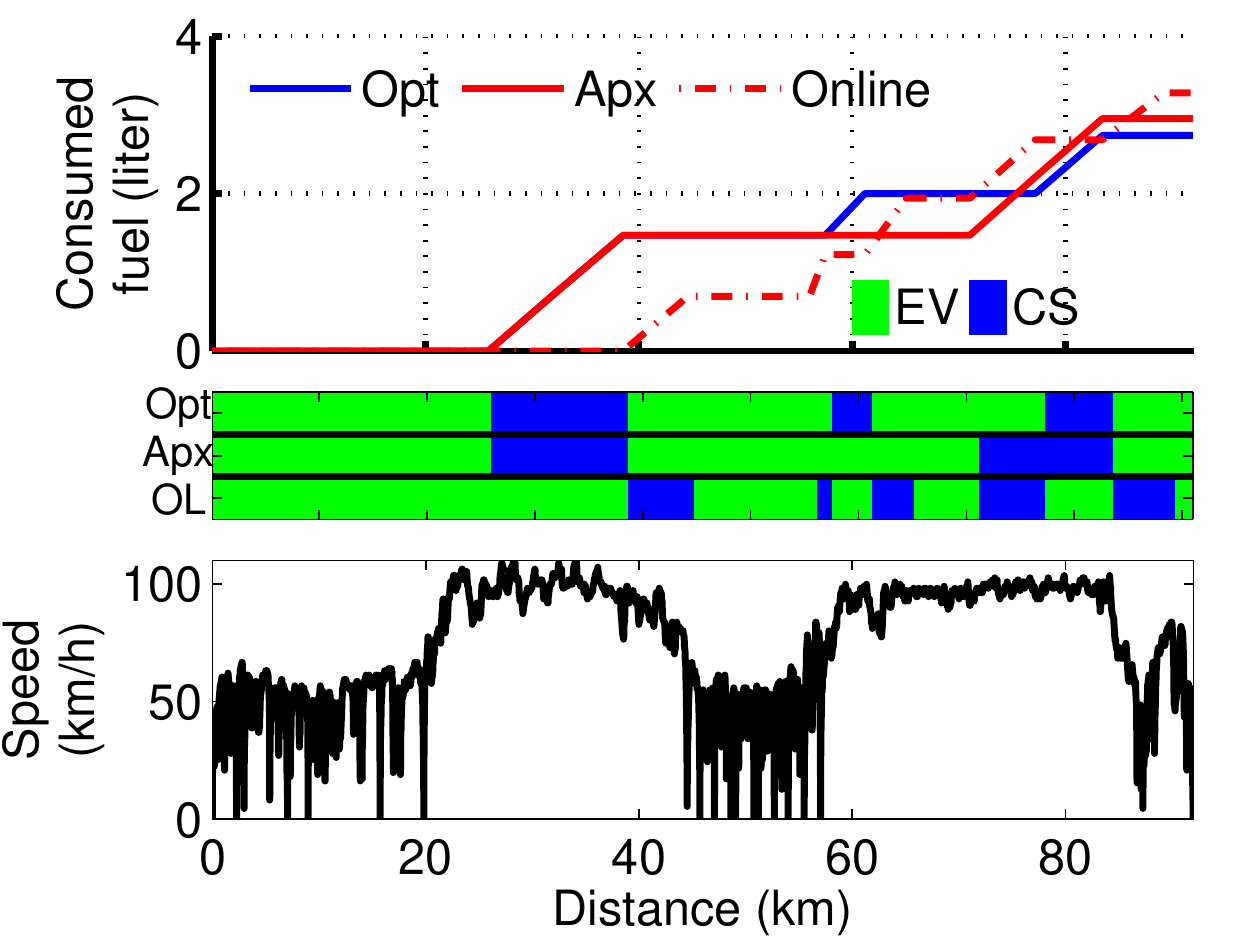}}
\subfigure[Case study of series/parallel hybrid on a route.]{\label{fig:opt2}\includegraphics[width=0.27\textwidth]{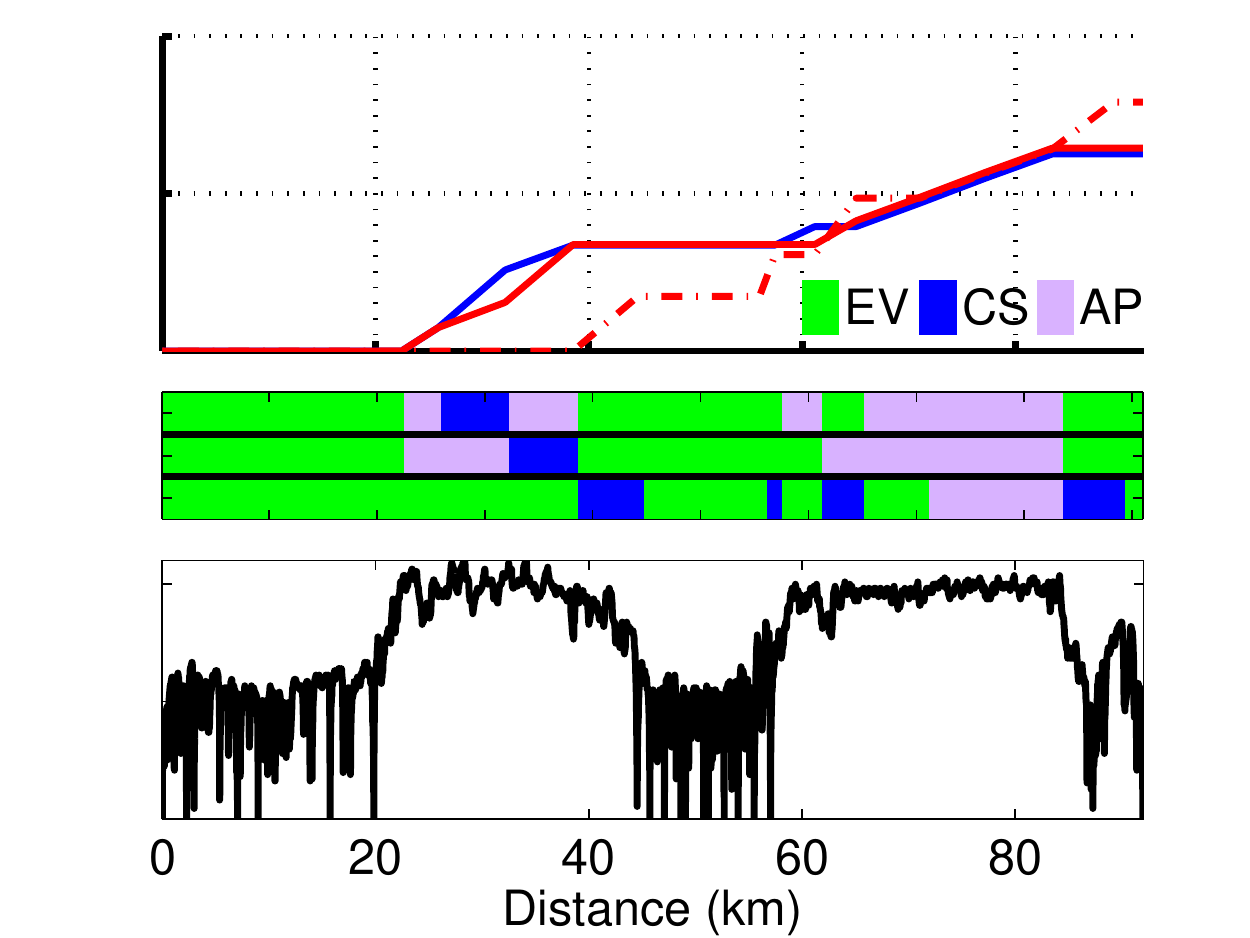}} \
\subfigure[Case study of series hybrid under different initial SoC.]{\label{fig:opt3}\includegraphics[width=0.22\textwidth]{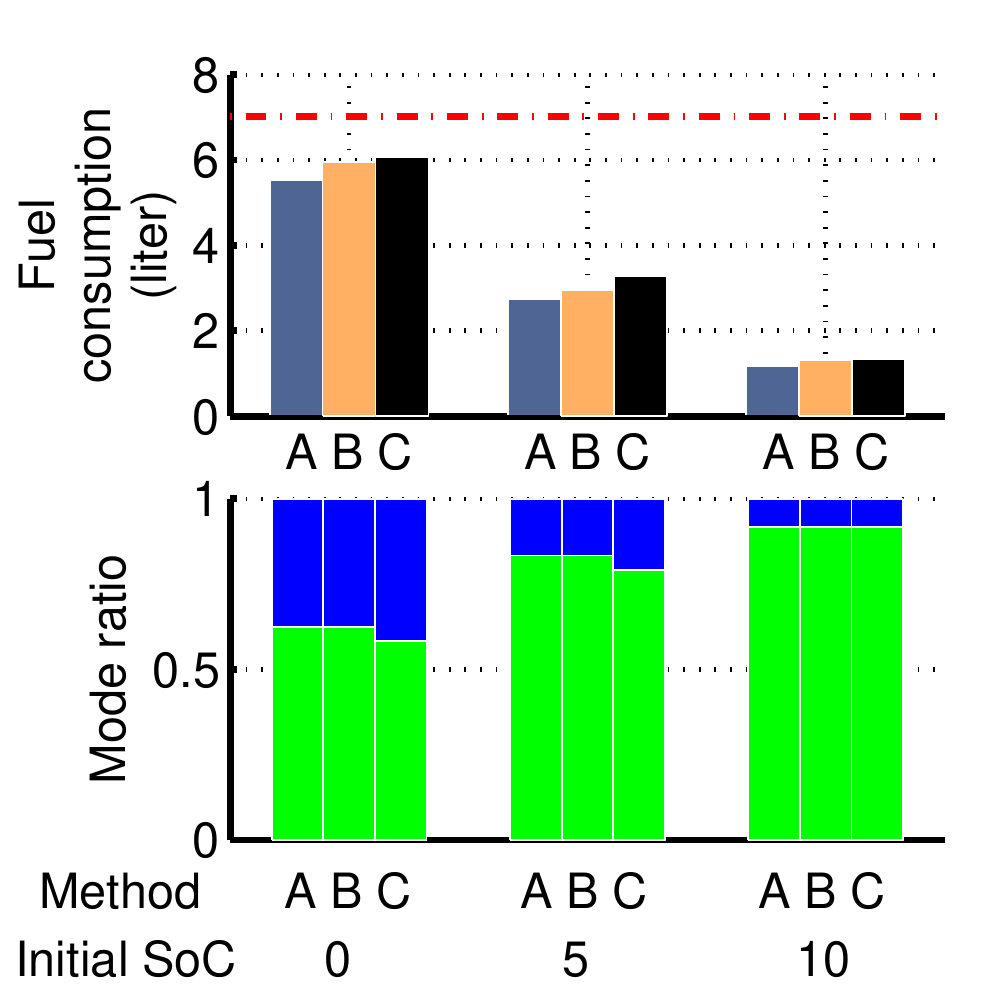}}
\subfigure[Case study of series/parallel hybrid under different initial SoC]{\label{fig:opt4}\includegraphics[width=0.22\textwidth]{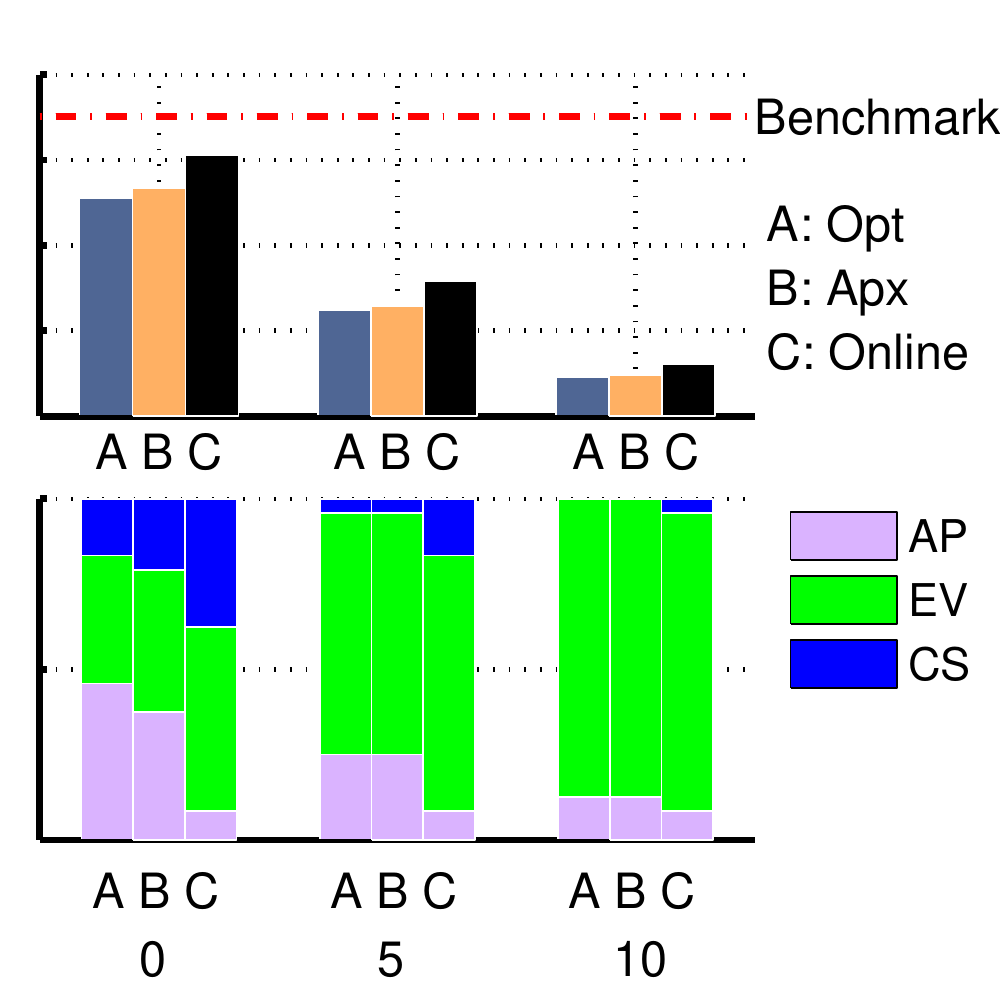}} 
\caption{Evaluation of route-based and online drive mode optimization algorithms.}
\end{figure*}

\begin{table}[!htb]
\caption{Parameters and values used in the model (Eqn.~(\ref{eqn:vehiclemodel}))} 
\centering{\scriptsize
  \begin{tabular}{@{}c|c|c|l@{}} 
\hline
\hline
    Parameters & Unit & Value & Description\\
\hline
$v$ & $m/s$ & - & velocity\\
$a$ & $m/s^2$ & - & acceleration or deceleration \\
${\sf m}$ & $kg$ & 1721 & car weight\\
${\sf g}$ & $m/s^2$ & 9.81 & gravity\\
$\rho_{a}$ & $kg/m^3$ & 1.226 & air density \\
$\alpha$ & $\circ$ & 0 & road grade\\
$A_f$ & $m^2$ & 2.202 & car frontal area\\
${\rm k}_d$ & - & 0.28 & car drag coefficient\\
${\rm k}_r$ & - & 0.01 & rolling resistance coefficient\\
\hline \hline
  \end{tabular}}
  \label{tab:paramdescript}
\end{table}

In Fig.~\ref{fig:modelfit}, we plot the estimation error, which is the difference between the accumulative measured energy consumption and the accumulative estimated energy consumption. We observe that the estimation error decreases substantially over a longer distance. This is because that the regression models are more accurate in capturing the long-term trends. We note that a more accurate model can be obtained by using more parameters, such as motor and engine RPM, or engine and motor efficiency maps. However, the simple yet rather accurate models we use are sufficient in practical application of drive mode optimization.

\begin{figure*}[htb!] 
\center
\setcounter{subfigure}{0}
\subfigure[Road network for case study.]
{\label{fig:map}\includegraphics[width=0.27\textwidth]{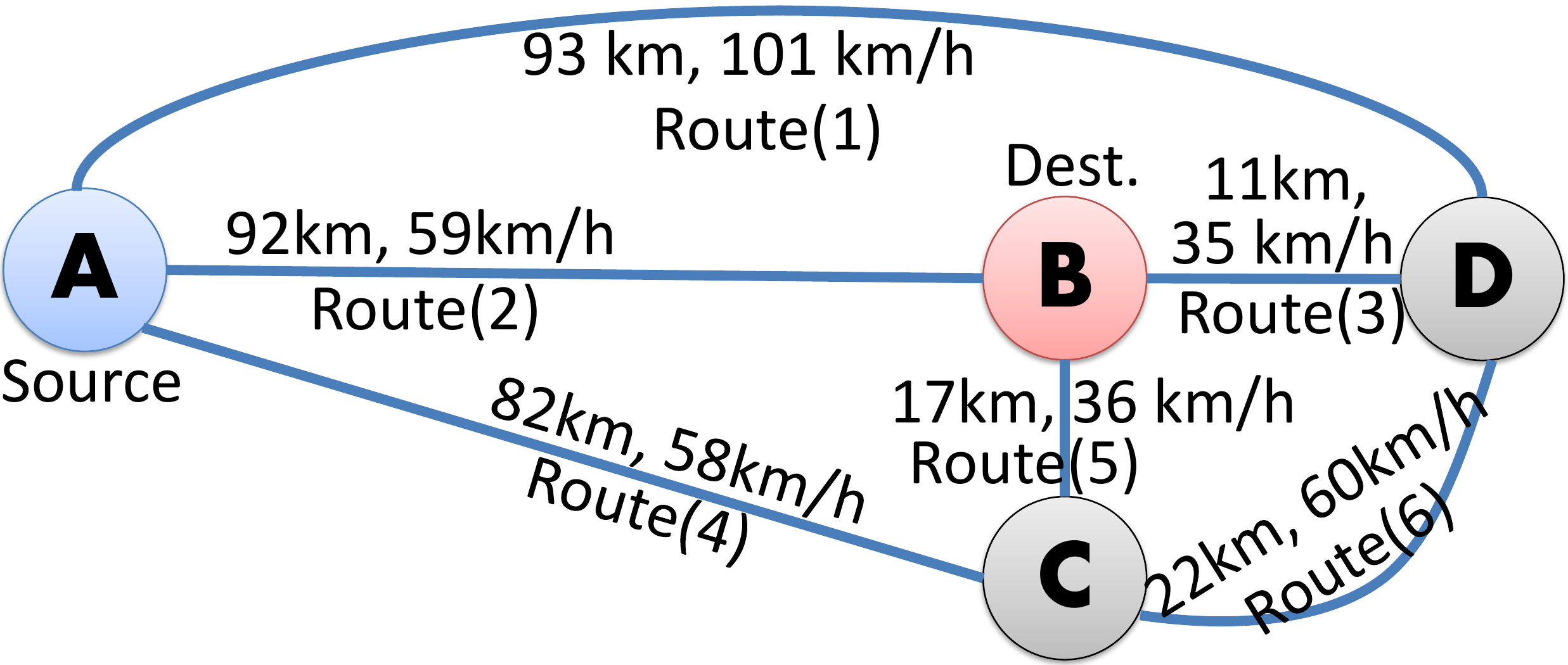}}
\subfigure[Case study of series hybrid.]{\label{fig:pathplanning1}\includegraphics[width=0.365\textwidth]{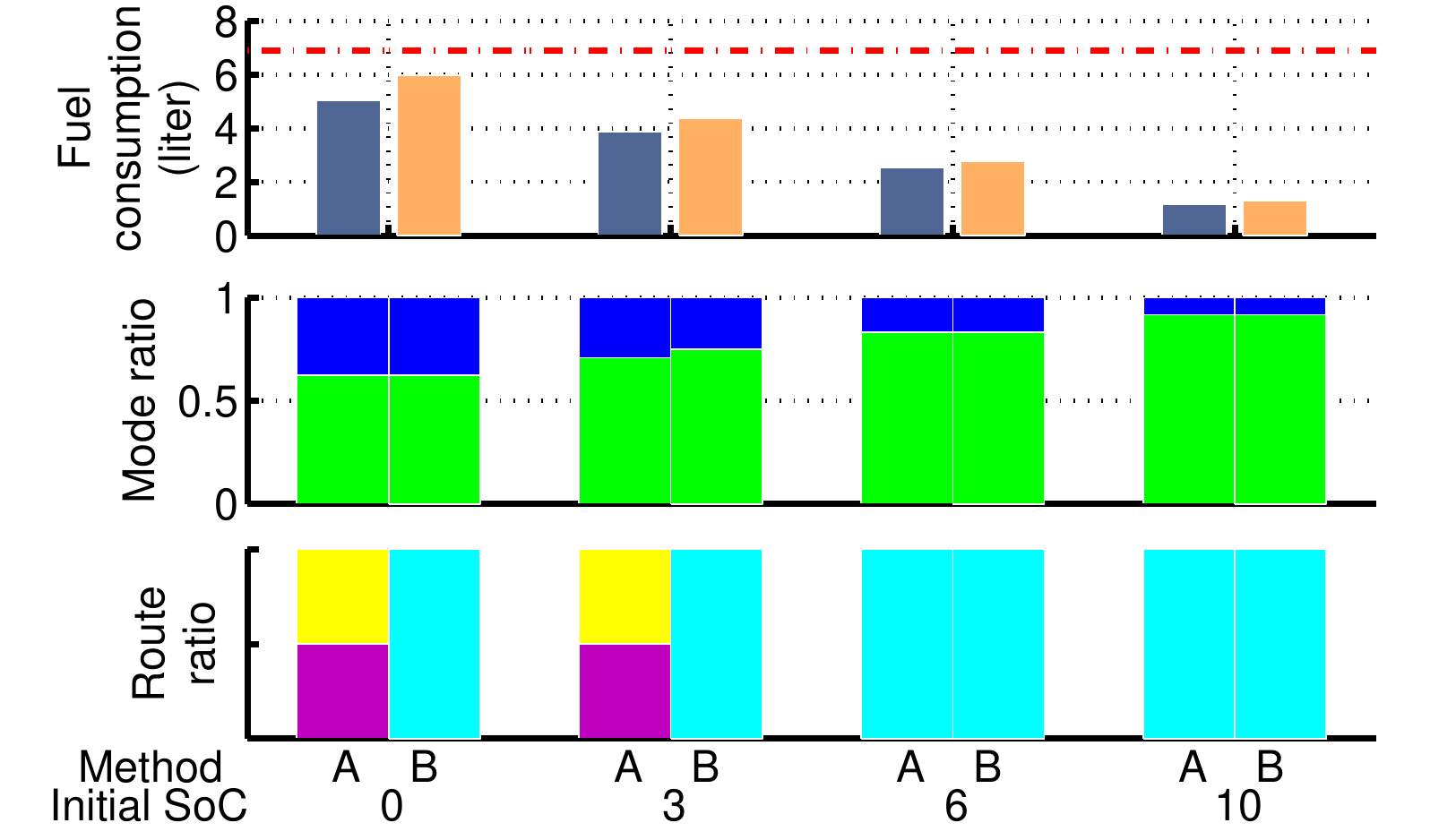}}\hspace{-1em}
\subfigure[Case study of series/parallel hybrid.]{\label{fig:pathplanning2}\includegraphics[width=0.365\textwidth]{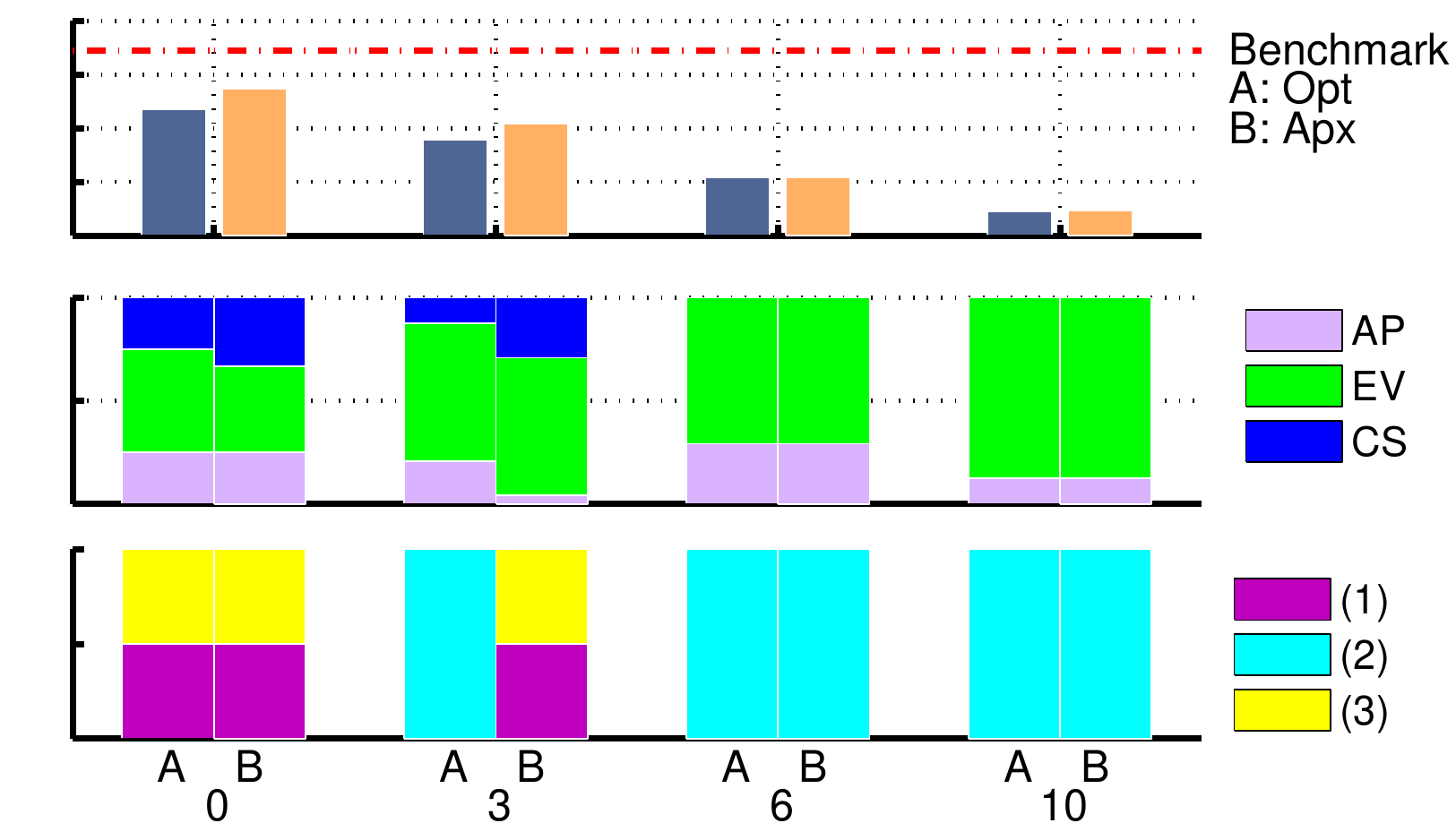}} 
\caption{Road network and the evaluation results for integrated path planning.}
\end{figure*}

\subsection{Drive Mode Optimization}

Using the estimated efficiency coefficients based on driving profile from the previous section, we compare the fuel consumption of our route-based and online drive mode optimization algorithms in two case studies: (1) series hybrid (with EV and CS modes) which simulates the Chevrolet Volt in the experiment, and (2) series/parallel hybrid (with EV, CS and AP modes) by augmenting an AP mode to the regression model of Chevrolet Volt and assuming $\beta_t=0.2$ in AP mode. We consider 3-minute time slots.
\begin{enumerate}[leftmargin=*]

\item{\em Drive Mode Decisions}: We collected the real-world driving profile on a specific route from an experiment with Chevrolet Volt in Figs.~\ref{fig:opt1}-\ref{fig:opt2}. We apply three algorithms (i.e., ${\tt Opt}$ using exact solution in Sec.~\ref{sec:dp}, ${\tt Apx}$ using approximation solution in Sec.~\ref{sec:relax}, and ${\tt Online}$ in Sec.~\ref{sec:online}) to the driving profile. The initial SoC is 5 kWh (i.e., half charged). We observe that ${\tt Apx}$ has a very similar fuel consumption with ${\tt Opt}$, whereas ${\tt Online}$ has slightly more fuel consumption than ${\tt Opt}$. We show the drive modes of each algorithm in Figs.~\ref{fig:opt1}-\ref{fig:opt2}. ${\tt Opt}$ and ${\tt Apx}$ give similar outcomes. Hence, ${\tt Apx}$ is a fast approximation to ${\tt Opt}$.

\item{\em Different Initial SoC}: Figs.~\ref{fig:opt3}-\ref{fig:opt4} show the fuel consumption under different initial SoC in the battery. To compare with the scenario without drive mode optimization, we consider a benchmark of using CS mode all the time. All ${\tt Opt}$, ${\tt Apx}$ and ${\tt Online}$ provide significant fuel savings, even in the presence of low initial SoC. Remarkably, without complete route information, ${\tt Online}$ can achieve considerable fuel savings, comparable to ${\tt Opt}$. Figs.~\ref{fig:opt3}-\ref{fig:opt4} also show the ratio of each drive mode. The more time using EV or AP, the more fuel savings can achieve.

\end{enumerate}
The algorithm is implemented in Matlab with 2.6 GHz Core i5, 8 GB RAM computer. The average running times of {\sc DMOP} are less than $10$ seconds, while the average running times of {\sc uPPDM} are around $100$ seconds.

\subsection{Integrated Path Planning}

Next, we evaluate our solutions on integrated path planning with drive mode optimization in two case studies: (1) series hybrid (with EV and CS modes), and (2) series/parallel hybrid (with EV, CS and AP modes). We collected driving profiles in a real-world road network, depicted in Fig.~\ref{fig:map} with four major stops, and the respective average speed and length of each route. Route(1) is a highway, whereas Routes(1)-(5) are regional roads. Node {\sf A} is the source, and node {\sf B} is the destination.

We apply two algorithms (i.e., ${\tt Opt}$ using exact solution in Sec.~\ref{sec:dp2} and ${\tt Apx}$ using approximation solution in Sec.~\ref{sec:relax2}) to the driving profiles on the road network. Figs.~\ref{fig:pathplanning1}-\ref{fig:pathplanning2} show the fuel consumption, drive mode ratio, route selections of each algorithm under different initial SoC in the battery. To compare with the scenario without path planning, we consider a benchmark of using CS mode all the time on the shortest route (i.e., Route(2)).

We observe that different optimal paths will be selected as a result of different initial SoC. For higher initial SoC (i.e., the battery is initially higher than half-full), Route(2) is the most energy-efficient, because there are more stop-and-go events for EV mode. For case study of series hybrid, ${\tt Opt}$ selects a different path when initial SoC is low, since it is most energy-efficient to take the highway (i.e., Route(1)) which can use CS mode to charge the battery for running EV mode later on Route(3) in the city.

The case study of series/parallel hybrid is similar to that of series hybrid. When initial SoC is low, the PHEV will select Route(1) to charge the battery and switch to EV mode on Route(3). We observe that series/parallel hybrid is more fuel-efficient than series hybrid on the same route. Also, series/parallel hybrid tends to use AP mode more frequently than CS mode on Route(2). Generally, the ratio of EV mode increase with higher initial SoC, because the PHEV requires less engine usage (e.g., AP mode or CS mode).

\section{Discussion and Conclusion} \label{sec:disc}

This paper investigated a driver-centric approach that enables the drivers to select the appropriate drive modes for minimizing fuel consumption. Optimization algorithms are presented to optimize drive mode decisions based on trip information, and integrated with path planning to consider intermediate filling and charging stations. An online competitive algorithm is provided that requires minimal a-priori trip information.  We implement our system and evaluate the results empirically on a Chevrolet Volt. We observe that our system can enable significant fuel savings.

We also address several practical issues of our approach.
\begin{itemize}

\item {\em Safety}: The drivers are supposed to switch their drive modes as informed by our system. Although this may present a distraction, our system can be adapted to only allowing to make drive mode decisions when the vehicle is stopped or moving at a safe speed. Also, we can increase the interval of time slots in the optimization problems to make less frequent drive mode decisions.

\item {\em ECO Mode}: ECO suppresses the vehicle performance, which is different from our approach that allows the vehicle to retain the desirable performance by optimizing drive modes accordingly. In certain PHEVs, ECO mode and other drive modes can be enabled simultaneously.

\item {\em Weather and Traffic}: Our system assumes mild weather and traffic conditions. We can extend our system to incorporate extra parameters to capture the impacts of weather in the vehicle model. Also, the vehicle speed can already reflect the expected traffic condition.


\end{itemize}
Recently, path planning problem is extended to aerial electric vehicles \cite{tcek2017drone}.

\bibliographystyle{IEEEtran}
\bibliography{reference}

\appendix

\begin{customthm}{1}
Algorithm ${\tt DMOP.DP}$ provides an optimal solution for \textsc{DMOP} with pseudo-polynomial running time. \end{customthm}
\begin{IEEEproof}
The basic idea of ${\tt DMOP.DP}$ is to enumerate every sub-problem $(\textsc{DMOP}[B_{t\mbox{-}1}, B_t, t])$ at each $t$ for all possible $(B_{t\mbox{-}1}, B_t)$. All the steps in ${\tt DMOP.DP}$ are evidently polynomial except those enumerating over the range of $B_t$.
Assume $\underline B$, $\overline B$, $B_0$, $\beta_t$, $\eta_{t}^{\rm r}$, $\eta_{t}^{\rm d}$, $\eta_{t}^{\rm e}$, $C_t$, $P_t^+$ and $P_t^-$ are given as rational number numerator at most $M\in\mathbb{Z}_+$ and common denominator $N\in\mathbb{Z}_+$. Then, Eqns.~\eqref{con:dmop-b}-\eqref{con:dmop-u} imply $B_t\in\left\{\frac{M'}{N^2}| M'\in\{0,1\ldots,TM^2\}\right\}$, which completes the proof.
\end{IEEEproof}

\smallskip

\begin{customthm}{2}
We consider the initial SoC $B_0 = \underline{B}$ and we require the final SoC to be $B_{T+1} = \overline{B}$.
Assuming $P^-_t = 0$ for all $t$, let the thresholds in Algorithm ${\tt Online}$ be $\theta^{\rm cs} = \sqrt{\frac{f_{\max}f_{\min}}{\kappa \eta^{\rm d}_{\min} \eta^{\rm e}_{\max}}}$ and $\theta^{\rm ap} = \theta^{\rm cs} \eta^{\rm d}_{\min} \eta^{\rm e}_{\min}$, where $\kappa \triangleq \max\{1,\frac{1}{\eta^{\rm e}_{\max}\eta^{\rm d}_{\min}} \}$,
then
${\tt CR}({\tt Online}) =  \sqrt{\frac{\kappa f_{\max} \eta^{\rm e}_{\max}}{f_{\min} \eta^{\rm d}_{\min}}} \frac{1}{\eta^{\rm e}_{\min}}$.
\end{customthm}

\begin{IEEEproof}
We consider the sequence $(P^+_t)_{t=1}^T$, such that each $P^+_t$ is satisfied by the respective drive mode. Since the initial SoC $B_0 = \underline{B}$, $P^+_t$ at any time $t$ must be satisfied by running CS mode before or at $t$. Also, since we require that the final SoC to be $B_{T+1} = \overline{B}$, always charging the battery up to $\overline{B}$ in ${\tt Online}$ will not incur unnecessary charging at the final $T$.

In ${\tt Online}$, the drive mode is selected according to priority: EV $\to$ AP $\to$ CS $\to$ CE, when the respective condition is met. For each $P^+_t$, let ${\sf Cost}[{\tt Online}, P^+_t]$ and ${\sf Cost}[{\tt Opt}, P^+_t]$ be the incurred cost by ${\tt Online}$ and offline optimal solution ${\tt Opt}$. Each $P^+_t$ can be satisfied by four cases in ${\tt Online}$:
\begin{enumerate}

\item[({\sf C1})] {\em Running EV mode at time $t$}: The battery has been charged by combustion engine in CS mode at some time slot before $t$ by ${\tt Online}$, which incurs a cost for $P^+_t$ at most $\theta^{\rm cs} \eta^{\rm d}_t \eta^{\rm e}_{\max} P^+_t$. But ${\tt Opt}$ incurs a cost at least $\min\{f_{\min} \eta^{\rm d}_t \eta^{\rm e}_{\min} P^+_t,  f_{\min} P^+_t \}$ (by running EV mode or AP mode at time $t$). The ratio of the cost of ${\tt Online}$ over ${\tt Opt}$ is upper bounded by 
\begin{equation}
\textstyle \frac{{\sf Cost}[{\tt Online}, P^+_t]}{{\sf Cost}[{\tt Opt}, P^+_t]}
\le  \frac{\theta^{\rm cs} \eta^{\rm e}_{\max}}{f_{\min}  \min\{\eta^{\rm e}_{\min}, \frac{1}{\eta^{\rm d}_{\max}}\}}
\end{equation}

\item[({\sf C2})] {\em Running AP mode at time $t$}: We consider the part powered by combustion engine (i.e., $\tilde{s}$). The incurred cost is at most $\theta^{\rm ap} \tilde{s}$. But ${\tt Opt}$ incurs a cost at least $f_{\min} \eta^{\rm d}_t \eta^{\rm e}_{\min} \tilde{s}$ (by running CS mode before $t$). The ratio of the cost of ${\tt Online}$ over ${\tt Opt}$ is upper bounded by 
\begin{equation}
\textstyle \frac{{\sf Cost}[{\tt Online}, P^+_t]}{{\sf Cost}[{\tt Opt}, P^+_t]}
\le  \frac{\theta^{\rm ap}}{f_{\min} \eta^{\rm d}_{\min} \eta^{\rm e}_{\min}}
\end{equation}
Note that the part powered by electric motor (i.e., $P^+_t - \tilde{s}$) can be addressed by ({\sf C1}).

\item[({\sf C3})] {\em Running CS mode at time $t$}: This implies that the battery has not been charged sufficiently by combustion engine in CS mode before $t$ by ${\tt Online}$. Otherwise, ${\tt Online}$ would run EV mode at $t$. The incurred cost for $P^+_t$ of ${\tt Online}$ at $t$ is at most $f_{\max} P^+_t$. But ${\tt Opt}$ may run EV mode at $t$ by charging the battery sufficiently by combustion engine in CS mode before $t$ with a cost at least $\theta^{\rm cs} \eta^{\rm d}_t  \eta^{\rm e}_{\min} P^+_t$. Otherwise, ${\tt Online}$ would also charge the battery sufficiently before $t$. Or ${\tt Opt}$ may run AP mode at $t$, which incurs a cost at least $\theta^{\rm ap} P^+_t$. Otherwise, ${\tt Online}$ would also run AP mode at time $t$. The ratio of the cost of ${\tt Online}$ over ${\tt Opt}$ is upper bounded by 
\begin{equation}
\textstyle  \frac{{\sf Cost}[{\tt Online}, P^+_t]}{{\sf Cost}[{\tt Opt}, P^+_t]}
\le  \frac{f_{\max}}{\min\{\theta^{\rm cs} \eta^{\rm d}_{\min} \eta^{\rm e}_{\min}, \theta^{\rm ap}\}} 
\end{equation}

\item[({\sf C4})] {\em Running CE mode at time $t$}: This case can be reduced to ({\sf C1}), because the cost of running CE mode is higher than AP mode, and $F(\cdot)$ is convex.

\end{enumerate}

Thus, the competitive ratio of ${\tt Online}$ is upper bounded by 
\begin{align}
& {\tt CR}({\tt Online})  =  \textstyle  \max_{\sigma} \frac{\sum_{t=1}^T{\sf Cost}[{\tt Online}, P^+_t]}{\sum_{t=1}^T{\sf Cost}[{\tt Opt}, P^+_t]} \\
 \le & \textstyle  \max_{\sigma, t} \frac{{\sf Cost}[{\tt Online}, P^+_t]}{{\sf Cost}[{\tt Opt}, P^+_t]}  \\
\le  & \textstyle \max \Big\{
 \frac{\theta^{\rm ap}}{f_{\min} \eta^{\rm d}_{\min} \eta^{\rm e}_{\min}},
\frac{\theta^{\rm cs} \eta^{\rm e}_{\max}}{f_{\min}  \min\{\eta^{\rm e}_{\min}, \frac{1}{\eta^{\rm d}_{\max}}\}},
\frac{f_{\max}}{\min\{\theta^{\rm cs} \eta^{\rm d}_{\min} \eta^{\rm e}_{\min}, \theta^{\rm ap}\}}
 \Big\} \notag 
\end{align}
An adversary will select the worst among the three cases.
We set $\theta^{\rm ap} = \theta^{\rm cs} \eta^{\rm d}_{\min} \eta^{\rm e}_{\min}$. Then
\begin{equation}
\textstyle \frac{\theta^{\rm ap}}{f_{\min} \eta^{\rm d}_{\min} \eta^{\rm e}_{\min}} \le
 \frac{\theta^{\rm cs} \eta^{\rm e}_{\max}}{f_{\min} \eta^{\rm e}_{\min}}
\end{equation}
In order to minimize the competitive ratio, we set 
\begin{equation}
\textstyle  \frac{\theta^{\rm cs} \eta^{\rm e}_{\max}}{f_{\min}  \min\{\eta^{\rm e}_{\min}, \frac{1}{\eta^{\rm d}_{\max}}\}} =
\frac{f_{\max}}{\theta^{\rm cs} \eta^{\rm d}_{\min} \eta^{\rm e}_{\min}}
\end{equation}
Let $\kappa \triangleq \max\{1,\frac{1}{\eta^{\rm e}_{\max}\eta^{\rm d}_{\min}} \}$. We obtain
\begin{equation}
\textstyle \frac{\theta^{\rm cs} \kappa \eta^{\rm e}_{\max}}{f_{\min} \eta^{\rm e}_{\min}}
= \frac{f_{\max}}{\theta^{\rm cs} \eta^{\rm d}_{\min} \eta^{\rm e}_{\min}}
\quad \Rightarrow \quad \theta^{\rm cs} = \sqrt{\frac{f_{\max}f_{\min}}{\kappa \eta^{\rm d}_{\min} \eta^{\rm e}_{\max}}} 
\end{equation}
Therefore, we obtain the competitive ratio as 
\begin{equation}
{\tt CR}({\tt Online}) = \textstyle  \sqrt{\frac{\kappa f_{\max} \eta^{\rm e}_{\max}}{f_{\min} \eta^{\rm d}_{\min}}} \frac{1}{\eta^{\rm e}_{\min}}
\end{equation} 
\end{IEEEproof}
\smallskip

\begin{customthm}{3}
Algorithm ${\tt PPDM.DP}$ computes an optimal solution for \textsc{PPDM} with pseudo-polynomial running time.
\end{customthm}
\begin{IEEEproof}
All the steps in ${\tt PPDM.DP}$ are evidently polynomial except that the dynamic program has to enumerate over the range of $B_t$.
By the same argument in the proof of Theorem~\ref{thm:t1}, this range is polynomial in the unary size of the input, and hence ${\tt PPDM.DP}$ is pseudo-polynomial.

To see that ${\tt PPDM.DP}$ is correct, we use the following extension of Lemma~2.1 in \cite{samir2011fillgas}:
\smallskip

\begin{lemma}\label{lem:l1}
In an optimal path (in $\tilde{\cal G}$), if ${\sf u}^{B}$ and ${\sf v}^{B'}$ are two consecutive nodes at which the vehicle stops for a fuel-refill, then the fuel level upon reaching ${\sf v}^{B'}$ must be either $0$, if $g_{\sf v}\le g_{\sf u}$, or $\overline G-w({\sf u}^{B},{\sf v}^{B'})$, if $g_{\sf v}> g_{\sf u}$. 	
\end{lemma}
\begin{IEEEproof}
If the conclusion of the lemma does not hold then the overall fuel cost can be reduced by an exchange argument. Indeed, if $g_{\sf v}\le g_{\sf u}$ but the fuel level upon reaching ${\sf v}_{B'}$ is strictly positive then reducing the fuel level at ${\sf u}$ by a (tiny) $\epsilon>0$ and increasing it at ${\sf v}$ by the same $\epsilon$ (while keeping the SoC the same) does not increase the overall cost and keeps the solution feasible. It follows that it is optimal to fill at ${\sf u}$ just enough to reach ${\sf v}$. Similarly, if $g_{\sf v}> g_{\sf u}$ and the fuel level at ${\sf u}$ is less than $\overline G$ then, since there is assumed to be a refill at ${\sf v}$, reducing the amount of refill at ${\sf v}$ by $\epsilon>0$ and increasing it at ${\sf u}$ by the same $\epsilon$ decreases the overall cost and keeps the solution feasible, which is contradiction ot the optimality of the initial solution.
\end{IEEEproof}
\smallskip

The above lemma justifies the definition of the set ${\mathscr G}({\sf v}^B)$ in Eqn.~\eqref{eq:e11}.
To see Eqn.~\eqref{recurrence1}, note that if the vehicle has to reach ${\sf t}$ form ${\sf u}$ in one hop, when the fuel level at ${\sf u}$ is $g$ and the SoC is $B$, then it has the option of recharging the battery up to $B'\in [B,\min\{B+E_{\sf u},\overline B\}]$, for a cost of $h_{\sf u}(B'-B)$ and refill the tank just enough to reach ${\sf t}$, for a cost of $(w({\sf u}_{B'},{\sf t})-g)g_{\sf u}$. Note that we use $w({\sf u}_{B'},{\sf t})$ which is the distance between ${\sf u}_{B'}$ and ${\sf t}$ in $\mathcal{G}_0$, and hence without any further stops for recharging; also $w({\sf u}_{B'},{\sf t})\le \overline {G}$ must hold, otherwise, it is impossible to drive from ${\sf u_{B'}}$ to ${\sf t}$ without refilling.

To see Eqn.~\eqref{recurrence2}, note that if the vehicle has to reach ${\sf t}$ form ${\sf u}$ in $q$ hops, when the fuel level at ${\sf u}$ is $g$ and the SoC is $B$, then according to Lemma~\ref{lem:l1}, it has the following options: (1) recharge the battery up to $B'\in [B,\min\{B+E_{\sf u},\overline B\}]$, for a cost of $h_{\sf u}(B'-B)$, and refill the tank just enough to reach the next stop ${\sf v}$ at an SoC $B''$, for a fuel cost of $(w({\sf u}^{B'},{\sf v^{B''}})-g)g_{\sf u}$; in this case we must have $g_{\sf v}\le g_{\sf u}$ by Lemma~\ref{lem:l1}, or
(2) recharge the battery upto $B'\in [B,\min\{B+E_{\sf u},\overline B\}]$, for a cost of $h_{\sf u}(B'-B)$, and fill up the tank, then reach the next stop ${\sf v}$, at SoC $B''$ and fuel level $\overline G-w({\sf u}^{B'},{\sf v}^{B''})$, for a fuel cost of $(\overline G-g)g_{\sf u}$; in this case we must have $g_{\sf v}> g_{\sf u}$. In both cases, the vehicle has to go from ${\sf v}^{B''}$ to ${\sf t}$ in $q-1$ hops.
\end{IEEEproof}

\end{document}